\documentclass[preprint2]{aastex}
\shorttitle{Formation of ultra compact dwarf galaxies}
\shortauthors{Tanuka Chattopadhyay et al.}

\begin{document}

\title{Uncovering the formation of ultra-compact dwarf galaxies
by multivariate statistical analysis }

\author{Tanuka Chattopadhyay}
\affil{Department of Applied Mathematics, Calcutta University,
92 A.P.C. Road, Calcutta 700009, India} \email{tanuka@iucaa.ernet.in}

\author{Margarita Sharina}
\affil{Special Astrophysical Observatory, Russian Academy of
Sciences, N. Arkhyz, KCh R, 369167, Russia}  \email{sme@sao.ru}

\author{Emmanuel Davoust}
\affil{IRAP, Universit\'e de Toulouse, CNRS, 14 Avenue Edouard Belin, 31400 Toulouse, France }
\email{davoust@ast.obs-mip.fr}

\author{Tuli De and Asis Kumar Chattopadhyay}
\affil{Department of Statistics, Calcutta University, 35 B. C.
Road, Calcutta 700019,India} \email {akcstat@caluniv.ac.in}

\begin{abstract}
We present a statistical analysis of the properties of a large
sample of dynamically hot old stellar systems, from globular
clusters to giant ellipticals, which was performed in order to
investigate the origin of ultra-compact dwarf galaxies.  The data
were mostly drawn from Forbes et al. (2008). We recalculated some
of the effective radii, computed mean surface brightnesses and
mass-to-light-ratios, estimated ages and metallicities. We
completed the sample with globular clusters of M31. We used a
multivariate statistical technique (K-Means clustering), together
with a new algorithm (Gap Statistics) for finding the optimum
number of homogeneous sub-groups in the sample, using a total of
six parameters (absolute magnitude, effective radius, virial
mass-to-light ratio, stellar mass-to-light ratio and metallicity).
We found six groups. FK1 and FK5 are composed of high- and
low-mass elliptical galaxies respectively. FK3 and FK6 are
composed of high-metallicity and low-metallicity objects,
respectively, and both include globular clusters and ultra-compact
dwarf galaxies.  Two very small groups, FK2 and FK4, are composed
of Local Group dwarf spheroidals. Our groups differ in their mean
masses and virial mass-to-light ratios. The relations between
these two parameters are also different for the various groups.
The probability density distributions of metallicity for the four
groups of galaxies is similar to that of the globular clusters and
UCDs. The brightest low-metallicity globular clusters and
ultra-compact dwarf galaxies tend to follow the mass-metallicity
relation like elliptical galaxies. The objects of FK3 are more
metal-rich per unit effective luminosity density than high-mass
ellipticals.
\end{abstract}


\section{Introduction}
The variety of astrophysical structures in the Universe, from
galaxies and galaxy clusters to stellar remnants, is well
described by essential physical principles \citep{P2000}. However,
their origin, in particular that of globular clusters (hereafter
GCs), is still a matter of debate. GCs are an intermediate cell of
structure between stars and galaxies, and  their formation process
is a cornerstone for our understanding the Universe
\citep{P1969,AZ92,h95,cot98}. If we consider star clusters as a
single class of astrophysical objects, there are many well-known
and poorly understood phenomena, which make their origin
enigmatic. For example, the color bimodality of GC systems
existing in most galaxies, the absence of a clear mass-metallicity
relation for the population of red GCs, the correlation between
color and integrated magnitude among the brighter metal-poor GCs
\citep{str08}, the differences in luminosity functions and surface
density profiles between young and old cluster systems
\citep[e.g.][and references therein]{bs06,l2010}. Spherical
stellar systems, whether globular clusters, elliptical galaxies,
or substructures of spiral galaxies, are considered virialized
\citep{Antonov73}. The origin of GCs is ultimately linked to the
evolution of larger pressure-supported structures within the
cosmological hierarchy \citep[e.g.][and references therein]{hw08}.

In the last decade a new type of
astronomical object has been discovered by a number of
astrophysicists \citep{hil99,ph2001,dri00,dri03,mie06}
while making a spectroscopic survey in
the Fornax cluster. These objects called ultra-compact dwarf
galaxies (UCDs), dwarf globular transition objects or
sometimes intermediate massive objects, are different from
the classical globular clusters or dwarf elliptical galaxies in
terms of their radii, relaxation time and V-band mass-to-light
ratios. They are more massive, more luminous, and have higher
mass-to-light ratio than globular clusters, but are fainter and
more compact than dwarf elliptical galaxies.

Several formation scenario have been proposed for understanding
their physical properties. \citet{kro98} and \citet{fel02}
suggested that UCDs are the results of merger of many young star
clusters formed in galaxy-galaxy encounters whereas \citet{mie02}
suggested that they are the luminous extension of massive GCs. The
formation of UCDs from the mass threshing of the envelopes of
nucleated galaxies has also been suggested
\citep{bas94,zin88,bek03,goe08} while along another line of
thought UCDs are considered fundamental building blocks of
galaxies \citep{dri04}. Special efforts were made to unite old
dynamically hot stellar systems, from GCs, UCDs and dwarf
spheroidals (dSphs) to giant elliptical galaxies
\citep{zar06,for08,dar08} to reveal the nature of UCDs.
\citep{mie08} have studied the internal dynamics of a large sample
of UCDs in Fornax. They argue that UCDs are dynamically unrelaxed
and dynamical evolution has probably not influenced their present
dynamical M/L ratio.

All these findings originated while studying two-point correlations
between different projections of the
fundamental plane of galaxies (FP) defined by velocity dispersion,
size (or effective radius) and surface brightness (or mass
density). For example, relations were found between size and
luminosity, mass and metallicity, mass-to-light ratio and
dynamical mass, luminosity and velocity dispersion etc.
Considering two parameters at a time means disregarding the
combining effects of others which in turn are responsible for
losing significant information.

For a unique and robust theory of the formation of UCDs, a
multivariate approach is more appropriate. The present work is
based on a data set covering a broad spectrum of objects,
including Galactic and extragalactic GCs, UCDs, young massive star
clusters, nuclei of dwarf ellipticals and pressure-supported
galaxies, presented in Section 2.  We have used the multivariate
statistical method of K-Means cluster analysis (presented in
Section 3), to classify these diverse objects with respect to a
set of physical parameters. Six homogeneous groups have been
identified by this objective method, they are described in Section
4, and their properties have been studied by several two-point
correlations and regressions (Section 5). Finally conclusions have
been drawn in Section 6.

\section{The data}

\begin{figure}
\epsscale{1.5}\plotone{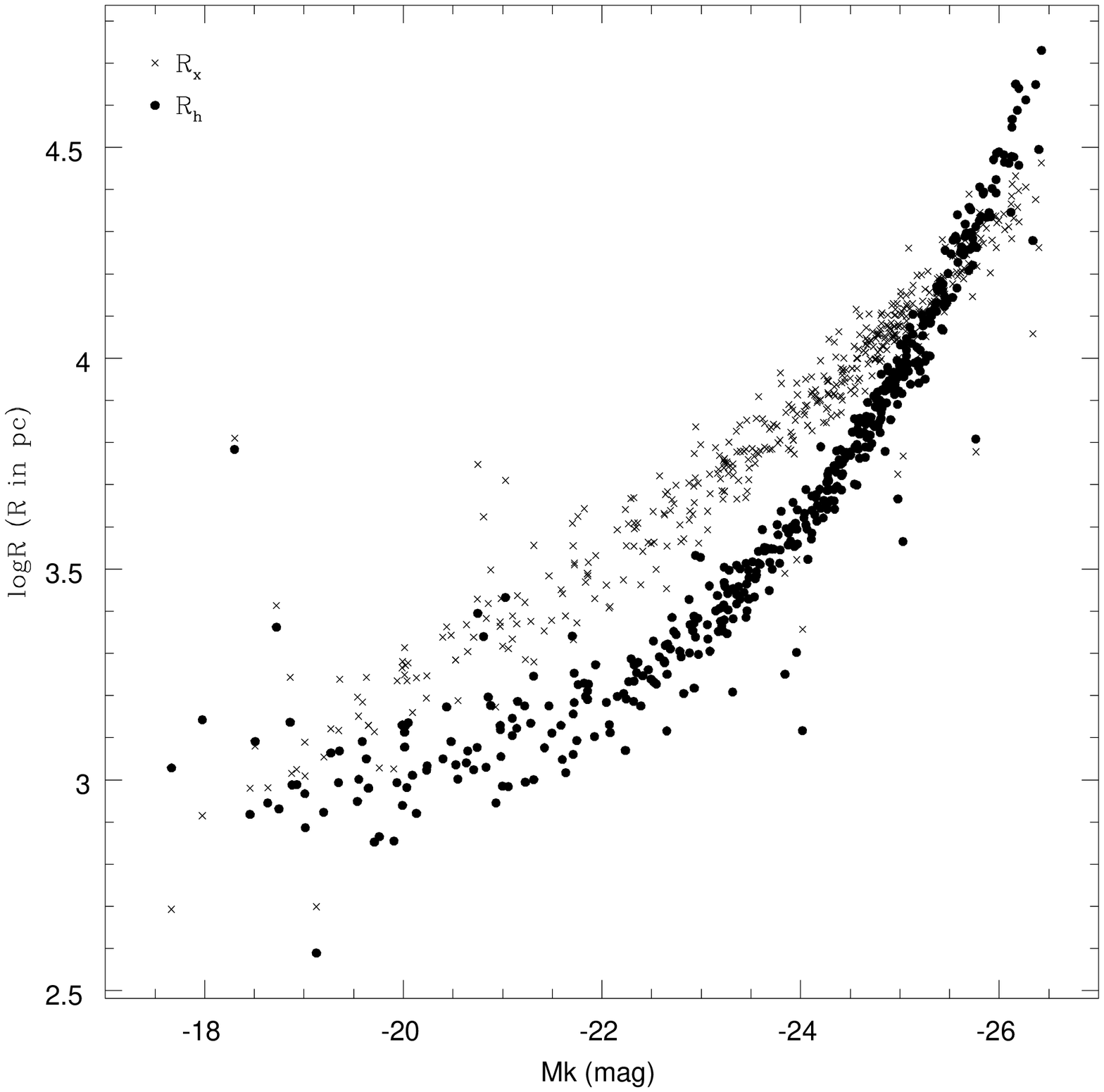}
\caption{Effective radius $R_h$
computed by us (filled circles) and $R_{20}$ from F08 (crosses)
versus absolute K magnitude $M_K$.  This figure is to be
compared to Fig.3 of F08.
\label{f1}}
\end{figure}

The present sample is composed of 370 objects from the paper of
\citet{for08}, hereafter F08, to which we added 19 GCs in M31. We
did not use all  objects of F08 because we were not able to
document all the values of the additional parameters (age,
metallicity, colors) used in the present study. We took the
distance, central velocity dispersion $\sigma_o$, effective radius
$R_h$ and apparent K magnitude $m_K$ from Table 1 of F08.  We did
not use the $R_h$ values of the galaxies given in Table 1 of F08
as they did not agree with Fig.3 of F08. Instead we recalculated
$R_h$ using $R_{20}$ following the method outlined in F08. To that
end we needed the axis ratio of the galaxies. We obtained $R_{25}$
from the hyperleda database\footnote{http://leda.univ-lyon1.fr}.
$logR_{25}$ is the logarithm of the axis ratio at the isophote 25
mag/$arcsec^2$ in the B band. It was available for all galaxies
except two (NGC1273 and PER195), which were removed from the
sample. The $R_h$ values that we obtained agree qualitatively with
those plotted in Fig.3 of F08 (see Fig.~\ref{f1}). To that sample
we added 19 GCs in M31. For these additional GCs the structural
parameters were taken from \citet{peo09}, the velocity dispersion
from \citet{str09} and the K magnitude from \citet{gal04}.
Hereafter we use the term IMO to designate dSphs and what F08 call
intermediate-mass objects, which include UCDs, young massive
stellar clusters, nuclei of dEs and M32.

We then derived the virial mass ($M_{vir}$) using the method
outlined in F08, the absolute K magnitude $M_K$ and the virial
mass-to-light ratio in the K band ($M_{vir}/L_K$).
We derived the effective luminosity density $I_e$ in the K band
(in $L_{K,\odot}/pc^2$) and the effective surface brightness
$\mu_{h,K}$ (in $mag/arcsec^2$) using the relations

\begin{equation}
Log I_e = 0.4(M_{K,\odot} - M_K) - log(2\pi) - 2logR_h
\end{equation}

\begin{equation}
\mu_{h,K} = M_{K,\odot} + 21.572 - 2.5 logI_e
\end{equation}

\noindent where $M_{K,\odot}$ = 3.28.

Next we derived or collected from the literature the metallicity,
broad-band colors, stellar mass-to-light ratio and age. The age is
the most poorly known parameter of all, except perhaps for the
Galactic GCs, whose relative ages are well known from studies of
deep color-magnitude diagrams \citep[e.g.][]{dea05,mar09}. The
term ``age" is defined precisely only for stars and globular
clusters, which originated in a single star forming burst. For
galaxies different methods give different age estimates.
Integrated characteristics, like colors, or narrow-band indices
are simultaneously influenced by metallicity and age, resulting in
a degeneracy. In this paper the term ``age" designates the age of
the main star formation period, and the metallicity of a galaxy is
its mean metallicity. Table 1 lists all the parameters considered
in the present work.

\subsection{Galactic GCs}

Metallicities in the \citet{zin84} scale were
extracted from the McMaster catalog \citep{har03}.
Ages in the \citet{zin84} scale were computed from the
corresponding relative ages in that scale extracted
from \citet{dea05}, \citet{mar09}, \citet{for10}.
$M/L_V$ were computed using the \citet{bru03}
models with the Padova (1994) tracks and the Chabrier IMF
and GC colors, corrected for Galactic extinction.

\subsection{GCs in M31}

Metallicities were (i) extracted from the catalog of \citet{gal09}
and (ii) calculated using a full set of Lick indices published by
\citet{puzia05} and \citet{b04,b05} and the program of
interpolation and chi-square minimization of \citet{sha06} and
\citet{sha09}. Ages were calculated using the full set of Lick
indices published by \citet{puzia05} and \citet{b04,b05} and the
program of interpolation and chi-square minimization of
\citet{sha06} and \citet{sha09}. For clusters without Lick indices
we used data from SED and fitting of \citet{wan10}. For the other
clusters only broad-band colors from \citet{gal04} and Mg2, Mgb,
Fe5270, and  Fe5335 from \citet{gal09} are available. We obtained
approximate ages using simple stellar population (SSP)
 models and the color/index data.
The latter ages are the least accurate.
$M/L_V$ were estimated using the \citet{bru03}
model dependence between age and $M/L_V$ at a
given age and $[Fe/H]$. \noindent All the derived data are in
agreement with the parameters published by \citet{c2011}.

\subsection{GCs in NGC5128}

Metallicities were taken from Table 4 of
\citet{dar08}. For GCs without metallicity
from the literature we calculated metallicities using a relation from \citet{sal07}:
$[Fe/H] = (3.87 \pm 0.07) (V-I) - (5.14\pm0.08)$.
Ages and $M/L_V$ were estimated
using broad-band colors from hyperleda, extinction data
and the \citet{bru03} models.
Internal extinction in NGC5128 was included.

\subsection{IMOs}

Age and metallicities of the young star clusters in the
remnant of the ``wet" merger NGC34 (W3, W30, G114) were taken from
\citet{sch07}. $M/L_V$ were calculated using SSP
models and photometric results presented in that paper. We used
broad-band photometry, metallicities, Lick indices and $M/L_V$
for VUCD 3, 4, 5 from \citet{evs07} to calculate their approximate age.
Evolutionary parameters and photometry results of the dwarf-globular
transition objects in the Virgo cluster (H8005, S314 to S999) were
adopted from \citet{hae05}. Photometry and metallicity
for the four UCDs in the Fornax cluster (UCD2 to 5) were
provided by \citet{mie06}. Ages, metallicities and $M/L_V
$ of the UCDs in the Fornax cluster (F-5 to
F59) were determined by \citet{chi11}.
The evolutionary parameters of the UCD M59cO
were taken from \citet{chi08}.
The metallicity for the dE VCC1254N was taken from
\citet{dur96}.

Colors and metallicities of dSphs in the Local Group were
taken from \citet{mat98}.  The peculiar
dwarf galaxy M32 was included in the list of IMOs (not Es) by F08
with morphological type cE (compact elliptical). We used the
data from \citet{mat98} for all its parameters, except
metallicity. Since M32 is seen through the disk of M31,
its very crowded surroundings make it
a complex case for photometry and spectroscopy.
We used the results of deep CMD
studies for the metallicity estimate \citep{gr1996,mon2011}
($[Fe/H]=-0.2$ dex). The difference between
the early \citep{dj92} ($[Fe/H]=-1.1 \pm 0.2$) and
later estimates is huge. However, a large spread of metallicities
and ages for stars has recently been found in M32 \citep{coe10,mon2011}.
The ages were taken equal to 13
Gyr for all dSphs, because all of them show ancient periods of star
formation according to CMD studies.

\subsection{Giant and dwarf elliptical galaxies}

We extracted $M/L_V$, ages and metallicities for 105 galaxies from
the literature \citep{J2004,Pr04,SB06,Li07,S08,A07,chi09}. These
all are spectroscopic determinations, except in the first paper.
For many of the sample ellipticals we have only broad-band colors
corrected for extinction from HyperLeda. We used SSP models of
\citet{bru03} and a Chabrier IMF to derive $M/L_V$, ages and
$[Fe/H]$ using broad-band integrated colors and magnitudes.

 Integrated colors not only depend on both mean
metallicity and age of a galaxy, but may also be affected by
internal extinction, possible ionized gas emission near the
galactic center, etc. There is a large number of unknown
parameters. To model the influence of age and metallicity on
integrated colors, we use the fundamental luminosity-metallicity
relation common for dwarf and giant ellipticals
\citep{Prugniel93,tho03}. We used the results of simulations of a
galaxy with exponentially declining star burst to derive the
approximate dependence of the broad-band colors on $M/L_V$, age,
and $[Fe/H]$ \citep{BJ01,mg05}. We selected colors more sensitive
to age or to metallicity. $I-K$ shows a minimal dependence on age
and $M/L_V$. There is a strong correlation between $B-R$ and
stellar M/L ratio independent of metallicity or star formation
rate \citep{BJ01}. $U-B$ is very sensitive to the age of a stellar
system and its M/L ratio. The slope of the color-magnitude
relation, and the color -- velocity dispersion ($\sigma$) relation
mainly depend on metallicity. Since mass is correlated with
$\sigma$ for Es, the color -- velocity dispersion ($\sigma$)
relation is equivalent to the mass metallicity relation. The age
difference between galaxies contributes mainly to the scatter of
the mass-metallicity relation. Fig.2 compares our metallicity
estimates for 105 ellipticals of our sample with values from the
literature.

\noindent Table 1 summarizes the data described in Sec. 2. The
successive columns give : name, log$R_h$ ($R_h$ in pc), $M_K$,
$\mu_{h,K}$, $M_{vir}/L_K$, $M/L_V$, the broad-band colors (U -
B), (B -V), (V - I) and (B - K), the metallicity ($[Fe/H]$) and
age determined by us, the metallicity and age from the literature,
the reference to the latter two data, and finally the group. Our
main contribution to the data is to have derived ages and
metallicities for 26 GCs in M31, ages and metallicities for Es,
and stellar mass-to-light ratios for most objects of the sample.

\section{The K-Means clustering technique}

Cluster analysis (CA) is the art of finding groups in
data. Over the last forty years different algorithms and softwares
have been developed for CA. The choice of a clustering
algorithm depends both on the type of data available
and on the particular purpose.

In the present study we have used the K-Means partitioning
algorithm \citep{mac67} for clustering. This algorithm constructs K clusters
i.e. it classifies the data into K groups which together satisfy
the requirement of a partition such that each group must
contain at least one object and each object must belong to exactly one group.
So there are at most as many groups as there are objects ($K \leq n$).
Two different clusters cannot have an object in common and the K groups
together add up to the full data set. Partitioning methods are
applied if one wants to classify the objects into K clusters
where K is fixed (which should be selected optimally). The aim
is usually to uncover a structure that is already present in the
data. K-Means is probably the most widely applied partitioning
clustering technique.

To perform K-Means clustering we used the MINITAB package. The
K-means clustering technique depends on the choice of initial
cluster centers. But this effect can be minimized if one chooses
the cluster centers through group average method \citep{mil80}. As
a result, the formation of the final groups will not depend
heavily on the initial choice and hence will remain almost the
same according to physical properties irrespective of initial
centers.

With this algorithm we first determine the
structures of sub populations (clusters) for varying numbers of
clusters taking K = 2, 3, 4, etc. Then using the Gap Statistics (see below)
we determine the optimum number of groups.

\subsection{The Gap Statistics}

In order to find the optimum number of groups we follow
the algorithm of Gap Statistics \citep{tib01}. Suppose
that a data set ${y_{il}}$, i = 1, 2, ..., n, l = 1, 2, ..., p, consists
of p features measured on n independent observations. Let $d_{ij}$
denote the distance between observations i and j. The squared
Euclidean distance $\sum_{l}$$(y_{il}-y_{jl})^2$ is used as a most
common choice for $d_{ij}$.
Suppose that the data have been grouped into k groups
G1, G2, ..., Gk, with Gr denoting the indices of observations in
group r, and $n_{r}$ is the number of observations in group r.
Let

\begin{equation}
D_{r} = \sum_{i,j\in G_{r}} d_{ij}
\end{equation}

\noindent be the sum of the pairwise distances for all points in
cluster r, and let

\begin{equation}
W_{k} = \sum \frac{1}{2n}D_{r}
\end{equation}

In the case that $d$ is the squared Euclidean
distance, $W_{k}$ will be the pooled within-cluster sum of
squares. The graph of $log(W_{_{k}})$ is standardized by
comparing it with its expectation under an appropriate null
reference distribution of the data. The estimate of the optimal
number of clusters is then the value of k for which $log(W_{k})$
falls the farthest below this reference curve. Hence
the gap is defined by

\begin{equation}
Gap_{n}(k) = E^{*}_{n}{log(W_{k})} - log(W_{k})
\end{equation}

\noindent where $E^{*}_{n}$ denotes the expectation from the reference
distribution. The estimate $\kappa$ will be the value maximizing
$Gap_{n}(k)$ on the basis of the corresponding sampling
distribution. As a motivation for the Gap Statistics, one may
consider clustering n uniform data points in p dimensions, with k
centers. Then assuming that the centers align themselves in an
equally spaced fashion, the expectation of
$log(W_{k})$ is approximately

\begin{equation}
log(pn/12)-(2/p)log(k) + constant
\end{equation}

In other words, the Gap Statistics is defined as the
difference between the log of the Residual Orthogonal Sum of
Squared Distances (denoted $log(W_{k})$) and its expected value
derived using bootstrapping under the null hypothesis that there
is only one cluster. In this implementation, the reference
distribution used for the bootstrapping is a random uniform
hypercube, transformed by the principal components of the
underlying data set. If the data actually have K well-separated
clusters, then it is expected that $log(W_{k})$ will decrease
faster than its expected rate (2/p)log(k) for $k\leq K$. When
$k \succ K$, then a cluster center is essentially added in the middle
of an approximately uniform cloud and simple algebra shows that
$log(W_{k})$ should decrease more slowly than its expected rate.
Hence the Gap Statistics should be largest when k = K.

\subsection{The algorithm to find the Gap Statistics}

Two common choices for the reference distribution are :
(a) each reference feature is generated uniformly over
the range of the observed values for that feature;
(b) the reference features are generated from a uniform
distribution over a box aligned with the principal components of
the data.

In other words, if X is an n$\times$p data matrix, it
is assumed that the columns have mean 0 and then the singular
value decomposition  X = UD$V^{T}$ is performed. It is transformed
through Y = XV and then uniform features, say T, are drawn over the
ranges of the columns of Y, as in method (a) above. Finally it is
back-transformed via Z = T$V^{T}$ to give reference data, say Z.
Method (a) has the advantage of simplicity. Method (b) takes into
account the shape of the data distribution and makes rotationally
invariant, as long as the clustering method itself is invariant.

In each case,  $E^{*}_{n}{log(W_{k})}$ is estimated by
an average of B copies $log(W^{*}_{k})$, each of which is computed
from a Monte Carlo sample $Y^{*}_{1} ,Y^{*}_{2} , ..., Y^{*}_{n}$
drawn from the chosen reference distribution. Finally, one
needs to access the sampling distribution of the Gap Statistics.
Let $s_{d}(k)$ denote the standard deviation of the B Monte Carlo
replicates $log(W^{*}_{k})$.  Accounting additionally for the
simulation error in $E^{*}_{n}{log(W_{k})}$ results in the
quantity $s_{k}$ = $\sqrt{(1+1/B)}s_{d}(k)$.
Using this the estimated cluster size $\kappa$ is chosen
to be the smallest k such that $Gap(k)\geq Gap(k+1)-s_{k+1}$.
where $s_{k+1}$ is a function of standard deviation of the
bootstrapped estimates.

The computation of the Gap Statistics proceeds as follows:
\begin{itemize}
\item Step 1: The observed data is clustered by varying the
total number of clusters from k= 1, 2, ..., K, giving
within-dispersion measures $W_{k}$, k = 1, 2, ..., K.
\item Step 2: B reference data sets are generated using
the uniform prescription (a) or (b) above and each one is
clustered giving within-dispersion measures
$W^{*}_{kb}$, b = 1, 2, ..., B, k = 1, 2, ..., K. Then the estimated
Gap Statistics is calculated as follows:
Gap(k) = $(1/B)\sum_{b}log(W^{*}_{kb})$-$log(W_{k})$.
\item Step 3. Let $\bar{l}=(1/B)\sum_{b}log(W^{*}_{kb})$, then
the standard deviation is computed as

$sd_{k}$=$[(1/B)\sum_{b}{log(W^{*}_{kb})-\bar{l}}^{2}]^{\frac{1}{2}}$

and $s_{k}$ is defined as $s_{k}$=$sd_{k}\sqrt{(1+1/B)}$.
\end{itemize}

Finally that number of clusters are chosen such that $\kappa$=smallest k and

\begin{equation}
Gap(k)\geq Gap(k+1)-s_{k+1}
\end{equation}

In other words the optimum number of clusters is that k
for which the difference

\begin{equation}
u_{k}=Gap(k)-(Gap(k+1)-s_{k+1}) \geq 0
\end{equation}

\section{Results}
\label{results}

The parameter set chosen for CA consists of $ M_K,
log(\sigma_0), logR_h, M_{vir}/L_K, [Fe/H], M/L_V$. Parameters
like $M_{vir}$, $\mu_{h,K}$ and age are not used. $M_K$

is very highly correlated with $M_{vir}$ and
$\mu_{h,K}$, so inclusion of these parameters does not influence
the clustering. Age is excluded because of the
large uncertainties associated to it.
The remaining parameters are not used because of the large number
of missing values. But, once the substructures are
identified, all the parameters are used to identify the distinctive
properties of the groups.

We have calculated the Gap Statistics for the set of six above
parameters, and the output suggests that the optimum number of
clusters is either four or six because the criterion used in Gap
Statistics to find the optimal number of clusters, i.e.
$u_{k}=Gap(k)-(Gap(k+1)-s_{k+1})\geq 0$ is  satisfied for k = 4
and k = 6.  Table 2 and Fig.3 show that the value of $u_{k}$
exceeds 0 for k = 4 and k = 6 and there is a sharp decline of the
graph after the value k = 6. Hence, considering all the criteria
discussed above, the optimal number of clusters for the present
sample is k = K = 6.  The six clusters (hereafter named groups to
avoid confusion with star clusters) are designated FK1 to FK6, and
their average properties are given in Table 2.

The elliptical galaxies were divided into two groups by the CA:
high-mass ellipticals (gEs) in FK1 and low-mass ones (dEs) in FK5.
Note that the labels ``gE" and ``dE" do not refer
strictly to the morphological types commonly used in astronomy. We use
these designations conditionally, to stress the statistical
difference in mass between the objects of FK1 and FK5.

FK3 has the high-metallicity GCs and the bright and
high-metallicity IMOs. The brightest
IMOs are UCD2 ($M_K=-16.32$), VUCD3 ($M_K=-16.21$), and UCD3
($M_K=-16.215$) \citep{hil07,evs07}.

FK6 is composed of IMOs, of the most massive GCs in the Galaxy,
in M31, and in NGC5128, (these GCs are all of low metallicity),
and of dSphs of the Local group : Leo~I
(Dist.$=0.25$ Mpc) and Sculptor (Dist.$=0.08$ Mpc).

Two groups, FK2 and FK4, have a negligibly small number
of members compared to the other groups : they contain three members each.
These are Local Group dSphs,
listed according to  their distances from the Sun in Mpc: UMi
(FK2, 0.066), Draco (FK2, 0.086), Sextans (FK2, 0.086), Carina
(FK4, 0.1), Fornax (FK4, 0.14), Leo II (FK4, 0.21).
We unfortunately do not have the full set of parameters for the other dSphs in the
Local Group and nearby groups to include them in the analysis.
A probable reason why these six objects were classified in such a way is their
M/L ratio, which is higher than for the other galaxies. These
two groups are considered only briefly, as their study
is the subject of a separate and elaborate study.  So there are
essentially four groups found as a result of our cluster analysis.

We show in Fig.4 how the various types of objects are distributed
among the different groups in $M_K$ - log$R_h$ space. The six
groups are indicated by different symbols, colored according to
the morphological type of the objects: GCs in black, IMOs in
green, and ellipticals in red.

To justify our choice of sample, and to show that it is
representative of stellar systems in the local Universe, we
compare it to that of \citep{mis11}, (hereafter MH2011) who, like
us and F08, studied a sample of stellar systems covering a large
range in masses, sizes and luminosities.

The sample of MH2011 is larger than ours, but the associated data
do not include velocity dispersions, metallicities or ages, so we
could not perform a similar analysis with their data.
Nevertheless, our sample covers basically the same space in
absolute magnitudes and effective radii, as shown in Fig.4, which
can be compared to Fig.1 of MH2011. The main difference is that
the sample of MH2011 has many more dEs (in the Hydra~I and
Centaurus clusters of galaxies), and extragalactic GCs (mostly GC
candidates in Virgo) and they included much fainter dwarf galaxies
of the Local Group, for which velocity dispersions would be very
difficult to measure. In short, our sample does not appear to be
biased against any particular type of object.

We also computed the probability density distribution (PDF) of
$M_K$ in our sample and compared it to the same distribution for
the MH2011 sample (see Fig.5). The method of non-parametric
density estimates is described in a previous paper \citep{cha09}.
The bin width for computing the density estimates is the same as
for the histograms shown in Fig.5.  Since MH2011 gives $M_V$
rather than $M_K$, we simply shifted their V magnitudes by 2.90,
which is the average value of (V - K) in our sample. There are
three main populations in both samples, the faintest one being
much more important in MH2011. Anticipating on our results, we
expect the distribution of metallicities for the F08 and MH2011
samples to be similar due to the fact that Es follow the
fundamental luminosity-metallicity relation
\citep{Prugniel93,tho03}.

The successive peaks are at $M_K$ = -24.8, -19.7, -13.3, -11 in
our sample, and at $M_K$ =  -25, -19.4, -14.8 in the sample of
MH2011.  The first peak in both samples corresponds to bulges and
the brightest elliptical galaxies. The next peak appears at the
location where the linear size - luminosity relation, common for
ellipticals and UCDs (MH2011), splits into two :  one relation for
dwarf galaxies and one for compact ellipticals and GCs. This
occurs at about $R_h$ = 1.3 kpc and $ Mass = 10^{10} M_{\odot}$.
So, galaxies in this group have roughly constant effective radii.
The faintest objects in this group have luminosities similar to
M32, $M_K \sim 18.5$, but their stellar densities are two orders
of magnitude lower (see Fig.5 in MH2011). The highest stellar
density for this group may be a characteristic scale, dividing
stellar systems into two systems. The internal acceleration for
one group is within the limits postulated in MONDian dynamics,
while for the other groups it is outside those limits. See also
the caption of Fig.7 of MH2011. The faintest broad PDF peaks
(-13.3 and -11 in our sample and -14.8 in MH2011) are different
for both samples. However, this is just a selection effect : as
mentioned above, our sample contains fewer dEs.

So, again, our sample does not differ significantly from another
large sample of stellar systems.
Our sample does not reflect the local luminosity
function for individual types of objects, and neither does the sample of
MH2011. We suggest that the relative intensity of the PDF peaks
in both samples reflects the way in which the samples were selected.

We also examined whether our choice of objects in the F08 sample
(370 out of 499) could bias the results in some way. We have
computed the mean $\pm$ standard error values of $M_K$ and
$log(\sigma _0)$ in the sub samples 1, 2, and 3 considered by F08
as well as for our corresponding sub samples. The number of
objects is of course different in the present sample and in the
F08 sample. But from Table 4 it is quite clear that this feature
does not introduce any significant bias as the mean values are
very similar.

We now present the distinctive properties of the groups, and look for
possible physical reasons for the differences and similarities between the
groups.

\section{Properties of the groups}

\subsection{Mass-to-luminosity ratios and binding energies}

We will discuss the virial M/L ratio, and it is important for what
follows to keep in mind that the stellar $M/L_V$ derived using
photometric data and SSP models is not necessarily identical to
the true baryonic M/L. This is due to the difficulty to correctly
take into account the star formation history (SFH) and initial
mass function of stellar populations \citep[e.g.][MH2011]{tra08}.
Furthermore, a disagreement between virial and baryonic M/L may be
due to the presence of dark matter, if the stellar population
model including SFH and initial mass function is correct.

The difference between virial and stellar M/L for our sample can
be seen from Table 1. It is seen that both the virial
($M_{vir}/L_K$) and the baryonic ($M/L_V$) mass-to-light ratios
differ at a high level of significance among the four main groups.
 Hereafter we will concentrate on $M_{vir}/L_K$ and
simply call it M/L. It is well known that UCDs tend to have higher
M/L than GCs \citep[][F08]{dar08}, and that dwarf spheroidal
galaxies have very high M/L from direct radial velocity
measurements of their brightest stars \citep[e.g.][]{sim07}.
Additionally, UCDs, like galaxies, have relaxation times greater
than the Hubble time \citep{kro98}. This is usually demonstrated
by plotting the data in the $k_1 - k_3$ space introduced by
\citet{ben92}, and this is well discussed in the aforementioned
papers.

For the present data, these parameters are :

\begin{displaymath}
k_1 = (log \sigma_0^2 + logR_h)/\sqrt{2}
\end{displaymath}

\begin{displaymath}
k_2 = (log \sigma_0^2 + 2 logI_e - logR_h)/\sqrt{6}
\end{displaymath}

\noindent
and

\begin{displaymath}
k_3 = (log \sigma_0^2 - logI_e - logR_h)/\sqrt{3}
\end{displaymath}

\noindent
where $I_e$ is given by Eq.~1.  These coordinates are simply
related to physical quantities : $k_1$ is proportional to the
logarithm of mass, $k_2$ is proportional to the effective surface
brightness times M/L, and $k_3$ is proportional to the
logarithm of M/L.

The differences in mass (represented by $k_1$) and M/L
(represented by $k_3$) between the groups are shown in Fig.6. The
groups occupy different locations in this projection of the FP,
except FK3 and FK6. For these two groups there is no continuity
break in the $k_1$, $k_3$ parameter distributions as for other
groups. Both FK3 and FK6 contain objects with high M/L. FK3
includes IMOs, and FK6 contains dSphs (Sculptor and Leo I) and
IMOs. We also note that the four main groups show wide and
different ranges in both mass and M/L. In each group, more massive
objects show higher M/L, but the slope of the correlation is
different for each group.

To quantify this, we performed robust multilinear regressions of
the form $k_3 = a + bk_1$ on the four main groups. The resulting
fits are listed in Table 5. The regression lines for the groups
FK3, FK5, and FK6 correspond to the relation M/L$ \propto$
M$^{0.2}$ within the errors. M/L is proportional to M$^{0.31}$ for
the group FK1. The position of the different objects within the
groups on the FP reflects not only differences in M/L, but also in
surface density, luminosity, and kinematical structure
\citep{djo87}. According to the slopes of the relations, the
objects in FK1 are much more influenced by the above three factors
than the objects in FK3, FK5 and FK6.

We now move on to discuss the edge-on projection of the
Fundamental Plane \citep{djo87,f76, kor77, d95} shown in Fig.7.
This figure is a representation of the Virial Theorem: $ r_e
\propto \sigma_0^2 I_e^{-1}(M/L)^{-1}$, usually applied to
galaxies \citep{f89,djo89}.  This figure also serves to compare
the binding energies of GCs \citep{mcl00}. The most compact and
luminous GCs have larger binding energies.

The groups FK3, FK5, and FK6 (i.e. GCs, IMOs and dEs) follow
roughly the same relation in the edge-on projection of the FP
(Fig.7). We obtained a bivariate least squares solution fitted
through $ \mu_{h,K}$:

\begin{displaymath}
log R_h-2 log\sigma_0=0.4(1.07(\pm0.03) \mu_{h,K}+19.1(\pm0.1))
\end{displaymath}

\noindent
which corresponds to $ r_e \propto \sigma_0^2 I_e^{-1.1}.$
The gEs of FK1 are concentrated in a parallel sequence, shifted towards lower
surface brightnesses ($ \mu_{h,K}$).  The bivariate correlation for FK1 gives:

\begin{displaymath}
log R_h - 2log\sigma_0 = 0.4(1.04(\pm 0.06) \mu_{h,K}+20.0(\pm0.13))
\end{displaymath}

These two solutions are close to the one that satisfies the Virial Theorem.
The different slope (1.07 in the first case, 1.04 in the second) is referred
to as the tilt in the FP, whose cause is still under debate
\citep[see][and references therein]{fr10}.
The tilt of the virial mass - total stellar mass relation common for
gEs, cEs and UCDs/GCs has been discussed in F08.
The difference in the zero points includes three components
\citep[e.g.][and references therein]{kor89}.
The first one reflects the density, luminosity and
kinematic structure of objects. The second factor indicates
whether the system is gravitationally bound or virialized.
If the deviation from the FP is due to mass-to-light ratio,
this implies the scaling relation $M/L \propto M^{0.2}$.
The systematic shift between gEs and the groups of GCs, IMOs,
and dEs is mainly due to the approximately ten times larger
M/L for gEs \citep{dar08}.

The objects of FK3 and FK6 are well mixed together in Fig.7, with
a tendency for FK3, which contains IMOs, to have higher binding
energy. The objects with the strongest deviation from the relation
are IMOs: e.g. B001, M59cO, UCD3; the globular cluster NGC2419,
and some dEs, like IC3779, with $\mu_{h,K} > 20$ mag
arcsec$^{-2}$. M32 has a very high binding energy, similar to that
of IMOs. Some gEs also fall in the same region of the diagram, as
M32, but no other dE does. \citet{bek01} and \citet{gra02} argued
that M32 is the stripped core of a larger galaxy. NCG2419 shows a
lower binding energy than other GCs. \citet{dar08} considered it
as the most likely candidate to host dark matter.

The $\mu_{h,K}$ versus $logR_h$ diagram (Fig.8) illustrates the
difference in stellar densities between the GCs of FK3 and FK6. It
shows that the GCs in FK3 have higher $\mu_{h,K}$ than those in
FK6 at a given $R_h$. In other words, FK6 has statistically
shallower surface brightness profiles than FK3. On the other hand,
the IMOs and GCs in FK3 are more massive/luminous and compact in
general than those in FK6.  \citet{jor05} found a significant
correlation between half-light radius and color for early-type
galaxies in the Virgo cluster in the sense that the red GCs are
smaller than the blue ones.

Having studied how mass is related to luminosity in our different groups,
we now examine how mass is related to metallicity.

\subsection{Mass-metallicity relation}

It is now well established that more massive galaxies are also
more metal rich; this is a consequence of the hierarchical
formation of galaxies in the Universe. But does such a relation
hold for all types of stellar systems?

\subsubsection{A boundary line}
The mass-metallicity relation (hereafter MMR) for our sample
objects is shown in  Fig.9a, where $k_1$, which is equivalent to
mass, is plotted versus $[Fe/H]$.

This figure shows that, except for a few objects, {\it all types
of stellar systems lie above a boundary line. It was plotted
 to stress the tendency, but} its slope
is surprisingly close to a MMR of the form $Z \propto Mass^{0.4}$.
The correlation is very weak for the objects in FK3 ($r(M_K,
[Fe/H]) = -0.382$) and FK6 ($r(M_K, [Fe/H]) = -0.177$), if we
consider them globally. Only the brightest low-metallicity GCs
(FK6) and IMOs at a given metallicity are close to the MMR. The
picture is almost the same if we plot absolute K magnitude versus
$[Fe/H]$ (Fig.9b). However, here the slope of the boundary line is
slightly different from that of the MMR: $[Z/H] \sim -3.5 - 0.14
M_K$.

What could be the origin of the boundary line?
It is unlikely to be caused by an observational selection effect.
We would presumably not see it if we included in the sample only
high-metallicity GCs and galaxies of other morphological types.
Many of the GCs are the brightest GCs of our Galaxy, and their
metallicities are very accurate. Extragalactic GCs and IMOs are also bright.
Their metallicities were obtained mainly via
spectroscopy, and are not very much influenced by observational errors and
the age-metallicity degeneracy. The only really uncertain metallicities are
those of ellipticals, because of the age-metallicity degeneracy and
uncertainties due to possible internal extinction, light-element
abundance variations, and large age and metallicity spreads within
individual galaxies. But, in spite of these uncertainties, the ellipticals
do follow the relation.

The slope of the boundary line is similar to that of the
luminosity-metallicity relation found in the literature. A
luminosity-metallicity relation,  $[Z/H]=-3.6-0.19 M_B$, was found
for dwarf and giant ellipticals in nearby galaxy clusters by
\citet{tho03}. It is equivalent to the equation $Z \propto
L^{0.4}$, found for dwarf galaxies in the Local Group by
\citet{dek86}, since $[Z/H]= [Fe/H] + 0.94 [\alpha/Fe]$
\citep{tho03}, $log (L_B/L_\odot) = 0.4(5.48 - M_B)$, and $log Z
\sim 0.977 [Fe/H] - 1.699$ \citep{b94}. We used here the solar
value $[\alpha/Fe]=0$. However, the deviations from this relation
for massive ellipticals may be large due to strong variations in
$[\alpha/Fe]$ for Es: $\sim 0.2 \div 0.5$ dex \citep[][and
references therein]{tho03,puzia06}. The median of the metallicity
distribution for elliptical galaxies and galactic bulges from the
Sloan Digital Sky Survey obtained by \citet{gal05} as a function
of stellar mass is also close to the relation $Z \propto M^{0.4}$
\citep[see also ][]{dar08}.

The origin of the luminosity-metallicity and mass-metallicity
relations for different morphological types of galaxies is still
an open issue \citep[e.g.][]{gre03,fd2008,ko00}. Does star
formation define the shape of the MMR? Does the boundary line mean
a lower fraction of matter  capable of being transformed into
stars under special physical conditions? It might result from the
interplay between internal and environmental factors: mergers and
interactions, inflows and outflows of gas, star formation
histories of individual galaxies in hierarchical galaxy formation.

The luminosity-metallicity relation for brightest GCs has been extensively studied
\citep{h06,mie06,P2009}. Using linear color-metallicity relations for {\it blue} GCs,
these studies derive scaling relations between GC luminosity L and metallicity Z
consistent with $Z \propto L^{0.5}$ \citep[e.g.][]{str08}.
The slope depends on the SSP models and on the light-element
abundances. According to \citet{C96} the mean $[\alpha/Fe]$ for Galactic GCs is 0.3 dex.
The same value was used by \citet{dar08} to calculate $[Z/H]$ for IMOs.

The metallicity of the faintest GCs close to the MMR is
intriguing. It corresponds approximately to extreme abundances of
Population II stars, i.e. stars formed immediately after the
initial pollution of interstellar medium by massive  Population
III stars: $Z \sim 0.01 Z_\odot$ \citep{Silk85}.

The Color-magnitude diagram (CMD) and chemical composition of some
GCs located near the border line (i.e. $\Omega$Cen, NGC2419, NGC
6341) are unusual. For example, NGC2419 is considered a remnant of
a dwarf galaxy due to its peculiar chemical composition
\citep{coh10}. The CMDs of most of these GCs show the existence of
multiple stellar populations, a fact that is still not fully
understood \citep[see e.g.][and references
therein]{bed2008,mar09,R08}. Since these GCs and IMOs are close to
the MMR, they were probably the brightest parts (nuclei) of
tidally destructed host galaxies \citep{zin88,bek03}. They may
also be geniune compact dwarf galaxies originating from
small-scale peaks in the primordial dark matter power spectrum
\citep{dri04}. GCs may have formed in dark matter minihaloes
\citep{mcs05}. However, it has not been established whether they
actually contain dark matter haloes  \citep{jor09,bau09}.

The Local Group dSphs and some GCs definitely fall below the
border line  in the $k1$ versus $[Fe/H]$ diagram. The reason has
been studied extensively for dSphs. Dwarf galaxies with
luminosities below some limit lose gas effectively because of
their low gravitational potentials, too shallow to prevent stellar
outflows following star formation episodes
\citep[e.g.][]{dek86,gre03}.

\subsubsection{Metallicity bimodality}

There is a gap between the groups FK6 and FK3 in Figs.9(a, b). It
is located near $[Fe/H] = - 1.0$ and is not horizontal.  The
low-metallicity peak is near $[Fe/H] = - 1.6 \pm 0.4$ dex, the
high-metallicity one is near $[Fe/H] = -0.6 \pm 0.04$ dex. Similar
metallicity peaks and the dividing line were identified for the GC
system of our Galaxy by \citep{h89}. GC systems of massive Es and
many spirals follow a bimodal color distribution
\citep{h96,gkp99,larsen01,P2006}.  A new feature shown in Fig.9 is
that the metallicity distributions of IMOs and extragalactic GCs
fall in the same range as the GCs of our Galaxy. Extragalactic
objects are gathered in two homogeneous groups together with
Galactic GCs.

Fig.10 shows the probability density distribution of $[Fe/H]$ for
all galaxies (FK1 + FK5 + FK4 + FK2), and for GCs and UCDs (FK6 +
FK3). The distribution is computed in the same way as the PDF
shown in Fig.5. Although the groups are not plotted with different
colors, they are clearly distinguishable from the PDF peaks. The
local maximum in the distribution for group FK3 at $[Fe/H] \sim
-0.6$ is close to that of FK5 ("dEs"), which is composed of
galaxies having a roughly constant effective radius and departing
from the size - mass relation common for gEs and UCDs (MH2011).
The corresponding PDF peak is also present in the probability
density distribution of luminosities (see discussion at the end of
Sec.4). The PDF for group FK6 corresponds to that of groups FK2+4,
i.e. galaxies less massive than 10$^8$ stellar masses.
Interestingly, there is another gap at the level of $[Fe/H]\sim
-0.3$, between six IMOs (M59cO, W3, W30, G114, VUCD 3, S490) and
the other objects in FK3. M32 with a central velocity dispersion
of $\sigma_0 \sim 79$ km/s also falls in this metallicity range.
However, since the luminosity and metallicity distributions of
galaxies are influenced by sample selection effect, the
correlation is not sufficient to establish the tidal origin of
nuclear GCs and UCDs.

The nature of the bimodality in the metallicity distribution is a
complex, still unanswered question. Due to the stochastic nature
of galaxy formation and star formation, hierarchical scenarios do
not reproduce the metallicity bimodality well. The dependence of
the galactic SFH on stellar mass is not straightforward
\citep{thomas05,R09}. Additionally, there is a strong
morphology-density relation, the environmental dependence between
stellar mass, structure, star formation and nuclear activity in
galaxies \citep[e.g.][]{K2004,R06}. Recent spectroscopic studies
have revealed strong age and metallicity gradients of different
slopes and values between the nuclear and outer regions of
elliptical galaxies \citep[][and references therein]{k11}. Nuclear
activity in galaxies often continues longer than in the outer
regions, because the fuel for star formation falls towards the
gravitational center. So, nuclei may contain multiple stellar
populations, and be on average younger and more metal-rich than
the rest of the galaxy. There are anomalous objects in both FK3
and FK6. In the low-metallicity group GCs like $\Omega$Cen,
NGC2419 and NGC 6341 show evidence of multiple stellar
populations.  The metal-rich GCs NGC6441 and NGC6338 have
prominent blue extensions in the horizontal branch \citep{ric97},
which are not typically associated with a globular cluster of this
metallicity, like 47 Tuc.

It has been proposed that the outer-halo GCs of the Galaxy were
accreted from the satellite galaxies
\citep[e.g.][]{mac04,cha07a,cha07b,mon08}. \citet{sh2010} suggest
that high-metallicity old GCs were formed from super star-forming
clumps with radii 1-3 kpc and masses $10^8$ to $10^9 M_\odot$,
which are known as a key component of star-forming galaxies at $z \sim 2$.\\

\subsubsection{Efficiency of metal production}

Is there a similar physical quantity for dynamically hot stellar
systems lying close to the MMR? For young stellar systems, for
example, {\it the star formation rate} is known to be an important
factor influencing the MMR  \citep[and references
therein]{man2010}. In Fig.11 we plot the dependence of $[Fe/H]$ on
the metallicity per unit effective luminosity density in the K
band. We call the last term {\it "metal production efficiency"}
(MPE) by analogy with the star formation efficiency (SFE), which
is the fraction of gas converted into stars at a particular
evolutionary stage of galaxies \citep{ken98}. MPE also reflects
the stellar density and the size of dynamically hot stellar
systems. Fig.11 shows that GCs and UCDs in FK6 have MPE in the
same range as Es: gEs (FK1) (MPE$ = 2.7 \pm 0.25$), and dEs (FK5)
(MPE$ = 1.7 \pm 0.4$). Galaxies with stellar masses $ M < 10^{10}
M_{\odot}$, including dSphs and GCs + UCDs (FK3+FK6), are in two
separate sequences, both showing a tendency for metallicity to
increase linearly with MPE. The two sequences intersect at the
location of the brightest UCDs and M32-like objects. Fig.11 also
shows that the objects of FK3 are the most metal-rich per unit
effective luminosity density. So, at least for these GCs and UCDs
in our sample, it is reasonable to assume that they are the
densest parts of galaxies accumulating fuel for star formation.

\section{Conclusion}

A multivariate statistical technique, K-Means clustering, has been
carried out on a data set taken from the paper of \citet{for08}.
It consists of elliptical galaxies, intermediate mass objects,
Local Group dwarf spheroidals, nuclei of dwarf ellipticals, young
massive objects and globular clusters. The sample properties were
completed by data from the literature or derived by us. Our aim
was to investigate the existence of interconnectedness, if any,
among the six groups found by our multivariate analysis.

In order to inquire into the  physical origin of IMOs, we
considered different projections of the 
fundamental plane using the results of the statistical analysis along with
observational data on velocity dispersion, effective radii and
effective surface brightness calculated from the total absolute
magnitude in the K band. We found that our groups are different in
terms of virial M/L ratios, and dependences between virial M/L ratios and mass.

The value of our study is that we include metallicities along with
other data in addition to the list of parameters of F08, which
definitely helps us to provide an objective classification into
groups. We consider a unified mass-metallicity dependence for all
the sample objects. It shows that  (i) there are GCs and UCDs in
the low-metallicity group sharing MMR with galaxies; (ii) there
are signatures of bimodality/multimodality in the metallicity
distribution that are common for GCs and IMOs on one hand, and for
low- and high-mass Es on the other hand. We speculate that the
rate of SF at the epoch when the objects were young is the
probable reason for the above two features. It appears that the
mean metallicities per effective K-band luminosity density (MPE)
for GCs and UCDs in FK6 lie in the same range as for elliptical
galaxies, suggesting similar physical processes and SFE. However,
MPE is much higher for GCs and UCDs in FK3. This confirms that
these objects originated as the densest parts of the present day
Es.

According to our findings, IMOs may be divided into two physical
groups: (i) Dwarf galaxy - globular cluster transition objects
formed in the same way and from the same material as old galaxies
and (ii) nuclei stripped from dwarf and normal ellipticals during
their dynamical evolution in groups and clusters. Note that since
UCDs were found only in dense environments, the last suggestion is
highly probable. Extensive theoretical and observational studies
are needed to establish the reasons for the described features and
the exact nature of UCDs.

\section{Acknowledgements}

T. C. thanks DST, India for supporting her a Major Research Grant.
M. S. acknowledges partial support of grants GK. 14.740.11.0901,
RFBR 11-02-90449 UKR-f-a, RFBG 11-02-00639-a, and thanks IRAP for
its hospitality. We thank the anonymous referee for detailed
comments which helped to improve the paper.


\clearpage

\begin{figure}

\includegraphics[angle=-90,scale=.60]{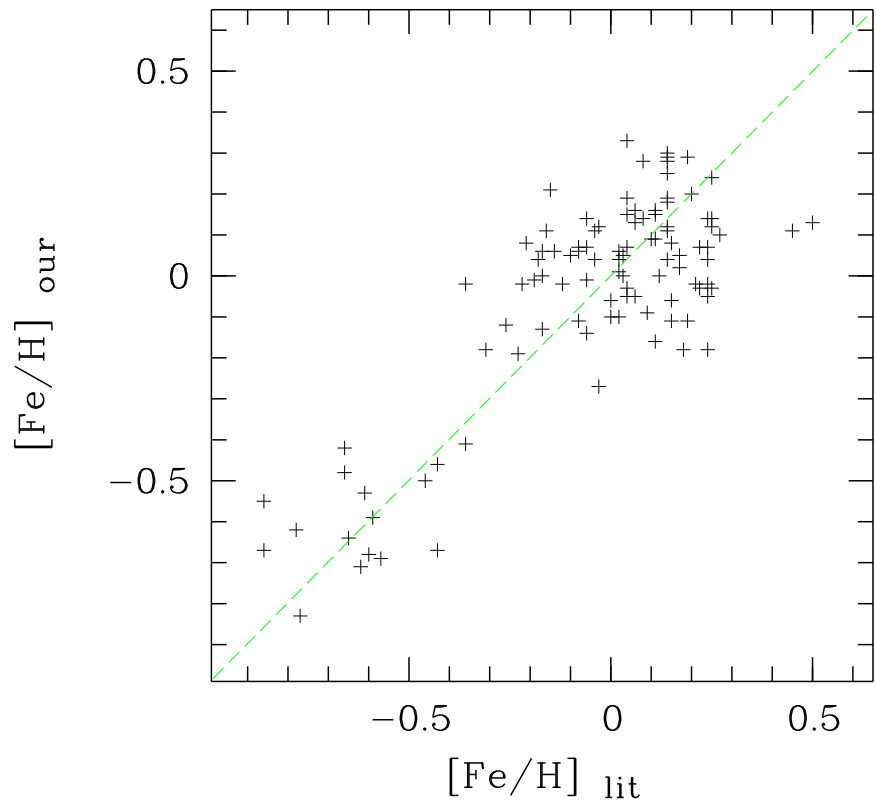}
\caption{Comparison of our derived metallicities of Es with values
from the literature (see Table~\ref{tbl_data})
\label{f2}}
\end{figure}

\clearpage

\begin{figure}
epsscale{2.0}\plotone{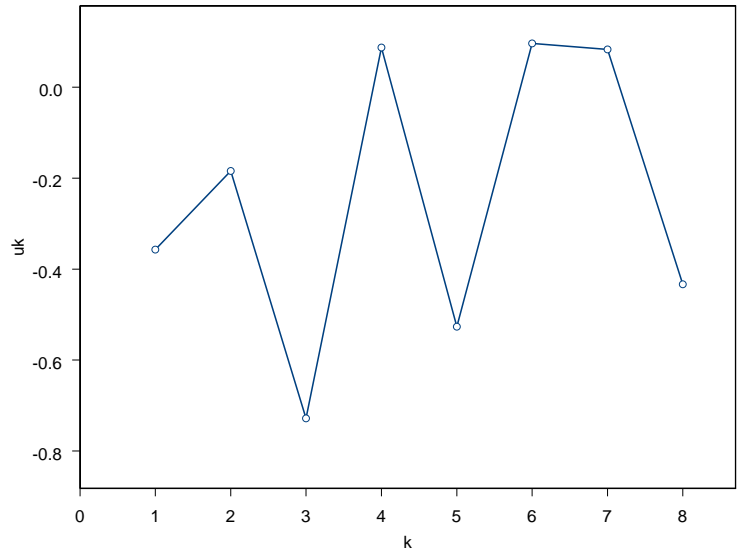}
\caption{The gap curve.
\label{f3}}
\end{figure}

\clearpage

\begin{figure}
\epsscale{1.5}\plotone{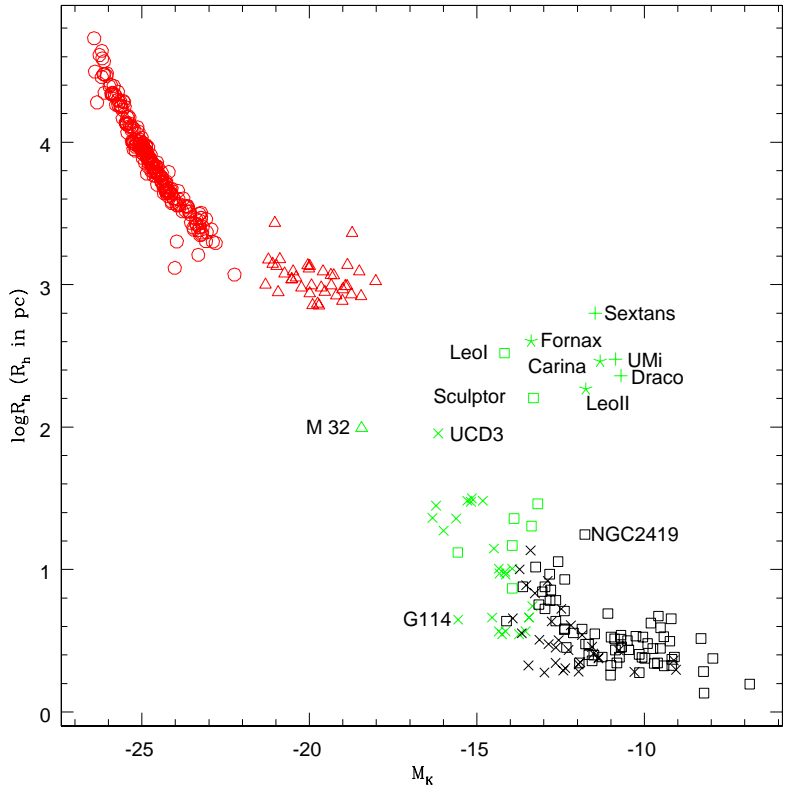} \caption{ Absolute K magnitude
versus logarithm of effective radius. The elliptical galaxies are
in red, the IMOs are in green and the GCs are in black. Open
circles are for FK1, pluses are for FK2, crosses are for FK3,
asterisks are for FK4, triangles are for FK5 and open squares are
for FK6. \label{f4}}
\end{figure}

\clearpage

\begin{figure}
\includegraphics[angle=0,scale=0.8]{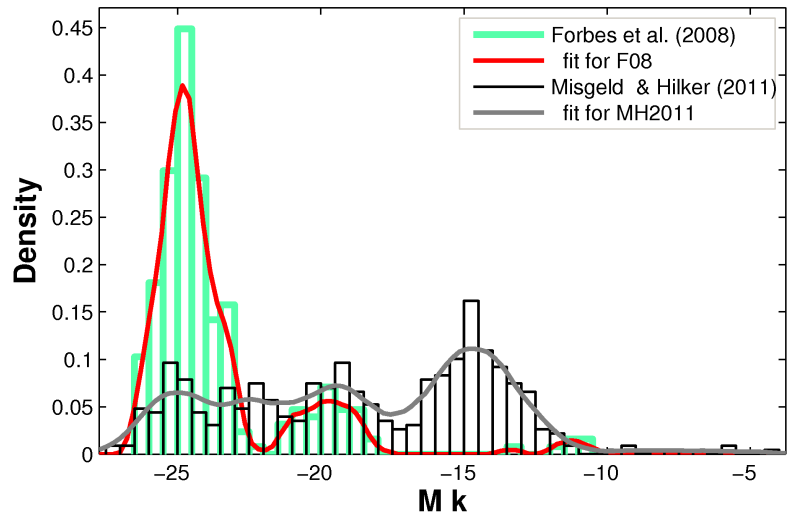}
\caption{ Probability density functions of absolute K
magnitude $M_K$. They are plotted in green for our sample and in
black for MH2011. The lines are non-parametric density
approximations. They are in red for our sample and in gray for
MH2011.
\label{f5}}
\end{figure}

\clearpage

\begin{figure}
\includegraphics[angle=-90,scale=.50]{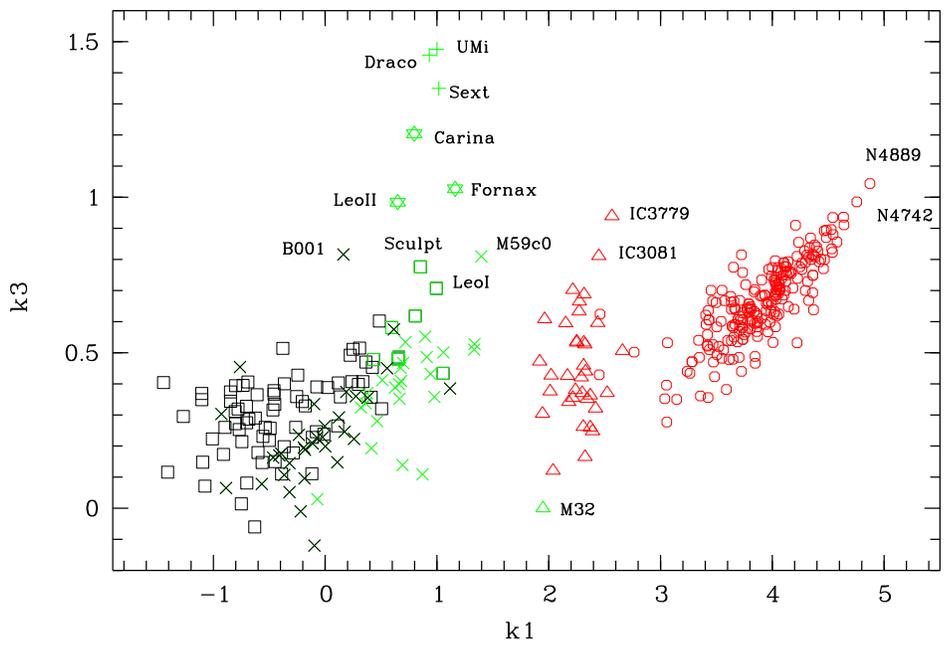}
\caption{Projection of the FP for dynamically hot stellar systems.
$k_1$ and $k_3$ are related to mass and M/L respectively. The
symbols and colors are the same as in Fig.\ref{f4}.
\label{f6}}
\end{figure}

\clearpage

\begin{figure}
\includegraphics[angle=-90,scale=.50]{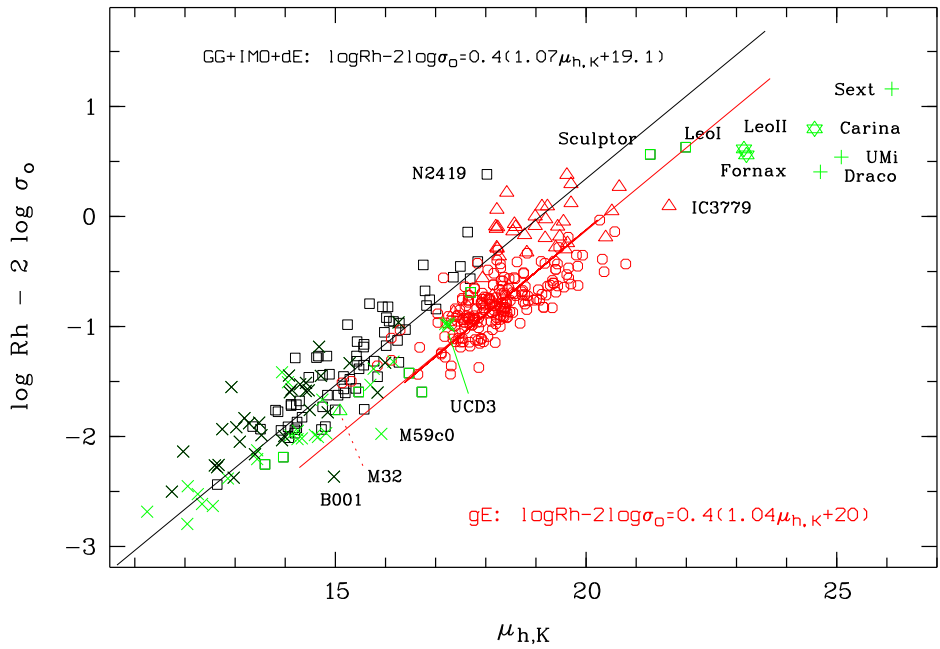}
\caption{Edge-on projection of the Fundamental Plane for the six
groups. The symbols and colors are the same as in Fig.\ref{f4}.
\label{f7}}
\end{figure}

\clearpage


\begin{figure}
\epsscale{1.5}\plotone{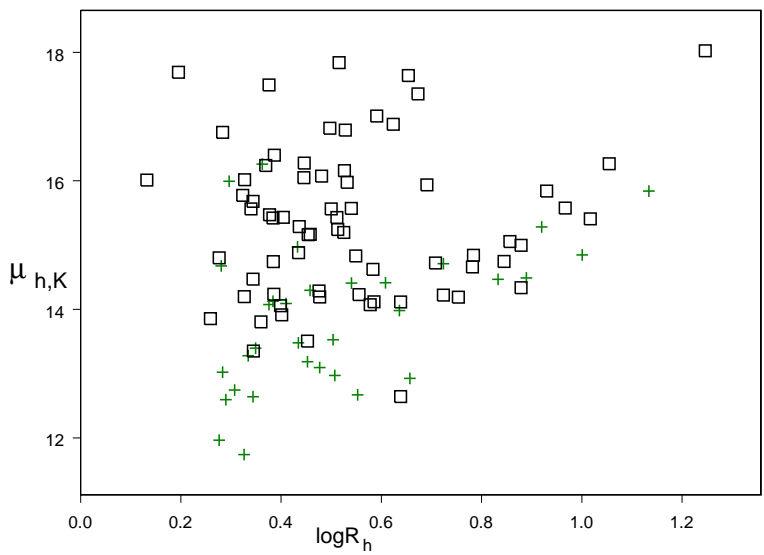}
\caption{The surface brightness
($\mu_{h,K}$) versus half light radius ($logR_h$) profile for GCs
in groups FK3 (green, plus) and FK6 (black,box) respectively.
\label{f8}}
\end{figure}

\clearpage

\begin{figure}
\includegraphics[angle=-90,scale=.45]{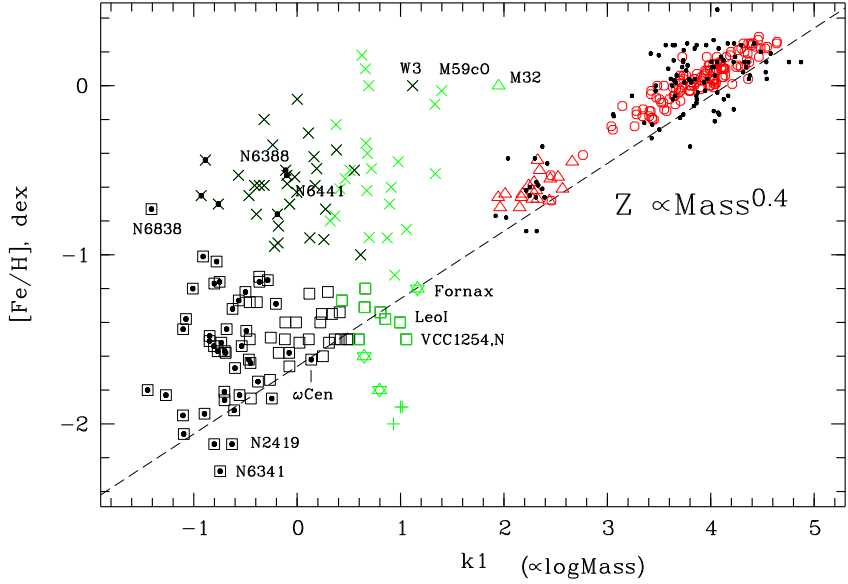}
\includegraphics[angle=-90,scale=.45]{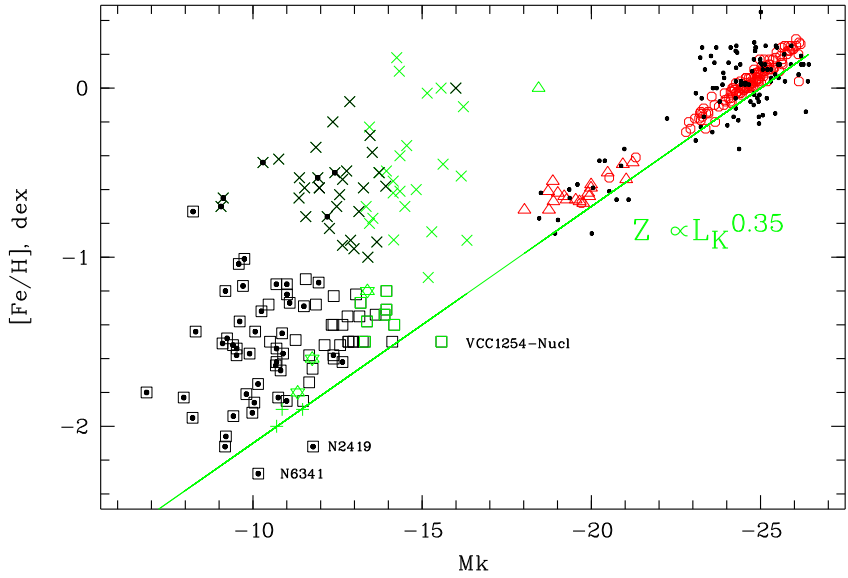}
\caption{ $k_1$ versus $[Fe/H]$ (a) and  $M_k$ versus $[Fe/H]$ (b)
for the six groups. The symbols and colors are the same as in
Fig.4. Black dots for ellipticals indicate metallicities from the
literature. Black dots for GCs (open squares) indicate Galactic
GCs \label{f9}}
\end{figure}

\clearpage

\begin{figure}
\includegraphics[angle=0,scale=0.80]{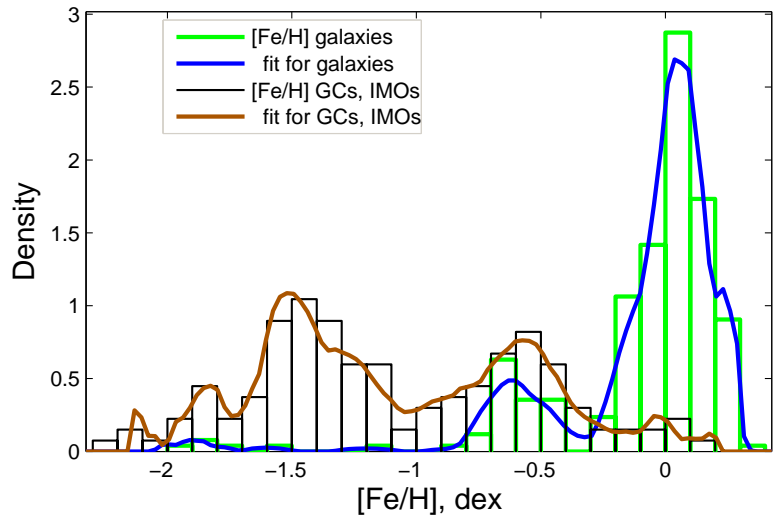}
\caption{ Probability density functions of metallicity $[Fe/H]$.
They are plotted in green for galaxies and in black for GCs and
IMOs. The lines are non-parametric density approximations. They
are in blue for galaxies and in brown for GCs and IMOs.
\label{f10}}
\end{figure}

\clearpage
\begin{figure}
\includegraphics[angle=-90,scale=.60]{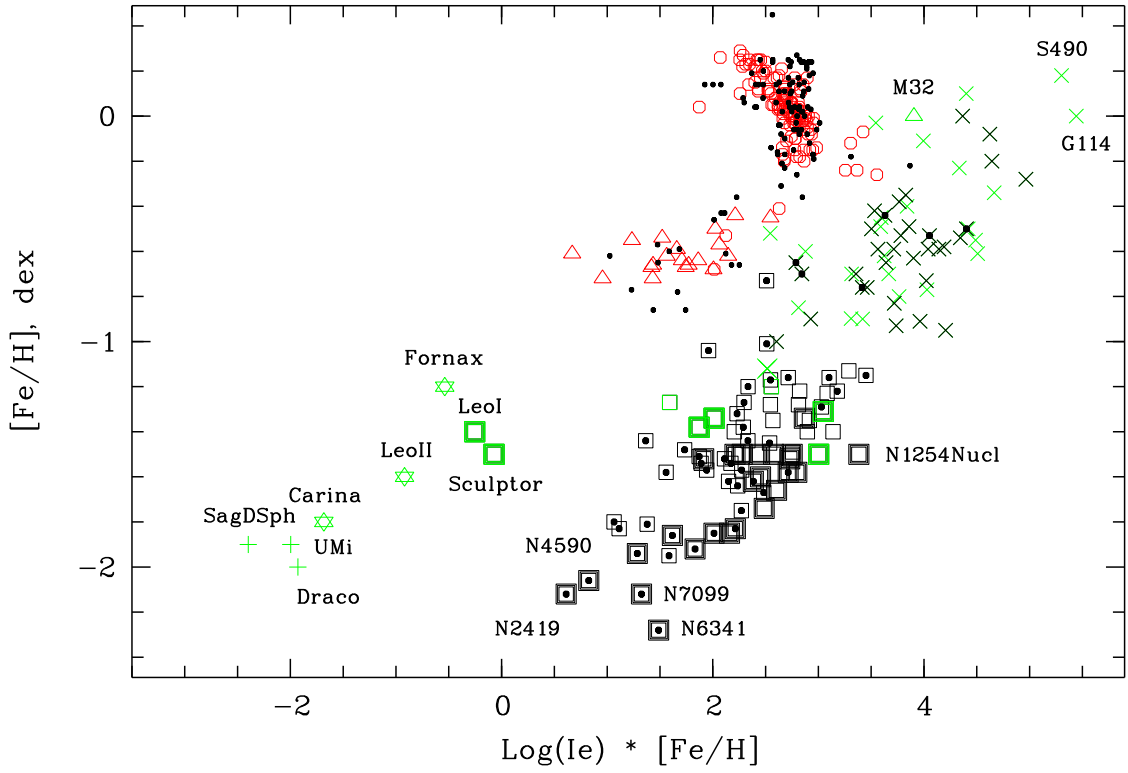}
\caption{$[Fe/H]$ versus metallicity per unit effective luminosity
density. The symbols and colors are the same as in Fig.9. Bold
symbols indicate that GCs and IMOs tend to follow the
mass-metallicity relation. \label{f11}}
\end{figure}

\clearpage

\begin{deluxetable}{lrrrrrrrrrrrrrlc}
\tabletypesize{\scriptsize} 
\rotate 
\tablecaption{Photometric and structural parameters.\label{tbl_data}}
\tablewidth{0pt} 
\tablehead{
\colhead{Name}&\colhead{logR$_h$}&
\colhead{M$_K$}&\colhead{$\mu_{h,K}$}&\colhead{M$_{vir}$/L$_K$}&
\colhead{M/L$_V$}&\colhead{(U-B)}&\colhead{(B-V)}&\colhead{(V-I)}&
\colhead{(B-K)}&\colhead{[Fe/H]}&\colhead{Age}&\colhead{[Fe/H]$_{lit}$}&\colhead{Age$_{lit}$}&
\colhead{Ref}&\colhead{Group}\\
\colhead{ }&\colhead{(pc)}&
\colhead{(mag)}&\colhead{(mag/arcsec$^2$)}&\colhead{(M$_{\odot}$/L$_{K,\odot}$)}&
\colhead{(M$_{\odot}$/L$_{V,\odot}$)}&\colhead{(mag)}&\colhead{(mag)}&\colhead{(mag)}&
\colhead{(mag)}&\colhead{ }&\colhead{(Gyr)}&\colhead{ }&\colhead{Gyr}&
\colhead{ }&\colhead{ }
}
\startdata
NGC104&0.60830&-12.19&14.41&0.851&2.70&0.33&0.84&1.09&3.63& \nodata & \nodata & -0.76 & 10.864& 9,10&3\\
NGC1851&0.25891&-11.01&13.86&0.931&6.36&0.15&0.74&0.98&3.42& \nodata & \nodata & -1.22 & 8.96& 9,10&6\\
NGC1904&0.48076&-9.90&16.07&1.074&1.85&0.05&0.64&0.90&2.68& \nodata & \nodata & -1.57 & 10.08& 9,10&6\\
NGC3201&0.59083&-9.51&17.01&1.973&1.80&0.12&0.73&0.91&2.85& \nodata & \nodata & -1.58 & 8.624& 9,10&6\\
NGC362&0.32645&-11.00&14.20&0.414&2.07&0.10&0.72&0.94&3.33& \nodata & \nodata & -1.16 & 18.512& 9,10&6\\
NGC4147&0.37595&-7.95&17.49&1.266&1.85&0.09&0.57&0.76&2.37& \nodata & \nodata & -1.83 & 11.536& 9,10&6\\
NGC4590&0.65406&-9.20&17.64&0.705&2.40&-0.02&0.58&0.87&2.45& \nodata & \nodata & -2.06 & 10.304& 9,10&6\\
NGC5272&0.51294&-10.89&15.24&0.539&3.30&0.08&0.68&0.92&2.64& \nodata & \nodata & -1.57 & 8.96& 9,10&6\\
NGC5286&0.34397&-10.82&14.47&0.797&1.85&0.01&0.64&0.86&2.92& \nodata & \nodata & -1.67 & 12.48& 9,10&6\\
NGC5694&0.52631&-10.04&16.16&1.172&2.40&-0.02&0.60&0.86&2.86& \nodata & \nodata & -1.86 & 11.76& 9,11&6\\
NGC5824&0.52518&-11.00&15.20&2.158&2.73&-0.03&0.62&0.87&2.81& \nodata & \nodata & -1.85 & 11.312& 9,10&6\\
NGC5904&0.69110&-11.09&15.94&0.703&1.33&0.14&0.69&0.91&2.97& \nodata & \nodata & -1.27 & 9.296& 9,10&6\\
NGC5946&0.34397&-9.61&15.68&0.520&1.50&-0.15&0.75&0.84&3.32& \nodata & \nodata & -1.38 & 10.08& 9,10&6\\
NGC6093&0.27664&-10.15&14.80&3.031&3.12&0.01&0.66&0.86&2.63& \nodata & \nodata & -1.75 & 10.864& 9,10&6\\
NGC6121&0.32705&-9.19&16.02&0.950&1.20&0.03&0.67&0.92&2.76& \nodata & \nodata & -1.2 & 10.192& 9,10&6\\
NGC6171&0.67324&-9.58&17.35&1.395&3.50&0.32&0.77&1.00&3.32& \nodata & \nodata & -1.04 & 10.976& 9,10&6\\
NGC6205&0.54000&-10.70&15.57&1.103&3.12&-0.04&0.66&0.83&2.66& \nodata & \nodata & -1.54 & 11.76& 9,10&6\\
NGC6218&0.49715&-9.23&16.82&1.541&1.85&-0.01&0.64&0.88&2.61& \nodata & \nodata & -1.48 & 10.528& 9,10&6\\
NGC6254&0.32346&-9.41&15.77&1.891&1.16&-0.08&0.62&0.83&2.63& \nodata & \nodata & -1.52 & 9.968& 9,10&6\\
NGC6256&0.29623&-9.06&15.99&2.388&0.51&-0.12&0.66&0.80&3.50& \nodata & \nodata & -0.7 & 13.5& 9,10&3\\
NGC6266&0.39873&-11.51&14.06&1.532&1.87&-0.01&0.72&0.93&3.18& \nodata & \nodata & -1.29 & 10.304& 9,10&6\\
NGC6284&0.53191&-10.25&15.98&1.244&1.44&0.09&0.71&0.92&3.07& \nodata & \nodata & -1.32 & 9.632& 9,10&6\\
NGC6293&0.37701&-9.98&15.47&1.676&1.39&-0.17&0.55&0.72&2.88& \nodata & \nodata & -1.92 & 13.5& 9,10&6\\
NGC6325&0.33994&-9.71&15.56&1.156&1.50&-0.15&0.77&0.93&3.79& \nodata & \nodata & -1.17 & 13.5& 9,10&6\\
NGC6341&0.40424&-10.16&15.43&0.916&1.80&-0.01&0.61&0.85&2.57& \nodata & \nodata & -2.28 & 12.096& 9,10&6\\
NGC6342&0.36245&-9.12&16.26&1.311&2.57&0.26&0.80&0.94&3.62& \nodata & \nodata & -0.65 & 10.304& 9,10&3\\
NGC6388&0.28980&-12.42&12.60&0.897&2.57&0.25&0.80&0.96&3.91& \nodata & \nodata & -0.5 & 10.8& 9,11&3\\
NGC6397&0.13211&-8.22&16.01&1.701&1.80&-0.08&0.55&0.78&2.19& \nodata & \nodata & -1.95 & 11.2& 9,10&6\\
NGC6441&0.34909&-11.92&13.40&1.484&2.57&0.28&0.80&0.97&3.22& \nodata & \nodata & -0.53 & 10.2& 9,10&3\\
NGC6522&0.38385&-10.07&15.42&1.225&1.40&0.11&0.73&0.85&3.27& \nodata & \nodata & -1.44 & 13.5& 9,10&6\\
NGC6535&0.19531&-6.86&17.69&1.957&0.95&-0.09&0.60&0.80&2.81& \nodata & \nodata & -1.8 & 10.44& 9,10&6\\
NGC6541&0.38437&-10.75&14.74&0.982&2.73&-0.03&0.62&0.82&3.04& \nodata & \nodata & -1.83 & 12.72& 9,10&6\\
NGC6558&0.51565&-8.31&17.84&1.572&3.60&0.09&0.67&0.86&2.65& \nodata & \nodata & -1.44 & 13.5& 9,10&6\\
NGC6624&0.28063&-10.30&14.68&0.507&2.50&0.29&0.83&1.03&3.72& \nodata & \nodata & -0.44 & 10.68& 9,11&3\\
NGC6626&0.43500&-10.86&14.88&1.091&3.30&0.01&0.68&0.86&3.48& \nodata & \nodata & -1.45 & 13.5& 9,10&6\\
NGC6656&0.45406&-10.68&15.16&1.481&1.85&-0.10&0.64&0.95&2.92& \nodata & \nodata & -1.64 & 12.36& 9,11&6\\
NGC6681&0.38645&-9.10&16.40&1.735&2.45&0.04&0.65&0.89&2.66& \nodata & \nodata & -1.51 & 10.416& 9,10&6\\
NGC6712&0.44554&-9.75&16.05&0.781&1.87&0.03&0.72&0.95&3.10& \nodata & \nodata & -1.01 & 10.4& 9,12&6\\
NGC6779&0.52818&-9.42&16.79&1.103&1.80&-0.07&0.66&0.88&2.76& \nodata & \nodata & -1.94 & 13.2& 9,11&6\\
NGC6809&0.62359&-9.80&16.88&1.446&4.62&0.02&0.64&0.89&2.92& \nodata & \nodata & -1.81 & 11.76& 9,10&6\\
NGC6838&0.28327&-8.23&16.75&0.621&2.70&0.35&0.84&1.02&3.55& \nodata & \nodata & -0.73 & 10.192& 9,10&6\\
NGC6864&0.45804&-10.69&15.17&1.929&1.44&0.10&0.71&0.94&2.90& \nodata & \nodata & -1.16 & 9.98& 9,12&6\\
NGC6934&0.44600&-9.52&16.28&1.352&3.60&0.09&0.67&0.85&2.76& \nodata & \nodata & -1.54 & 9.52& 9,10&6\\
NGC7089&0.51139&-10.70&15.43&1.377&0.95&0.02&0.60&0.84&2.29& \nodata & \nodata & -1.62 & 10.528& 9,10&6\\
NGC7099&0.36952&-9.18&16.24&1.879&1.85&0.00&0.57&0.82&2.32& \nodata & \nodata & -2.12 & 12.096& 9,10&6\\
NGC2419&1.24613&-11.78&18.02&0.300&1.39&-0.05&0.55&0.90&2.78& \nodata & \nodata & -2.12 & 12& 9,13&6\\
B001&0.43297&-10.76&14.97&10.113&2.30&\nodata&\nodata&\nodata&\nodata& 0.15 & 0.17 & \nodata & \nodata & \nodata &3\\
B006&0.54033&-11.86&14.41&1.000&2.10&\nodata&\nodata&\nodata&\nodata& -0.35 & 10.2 & \nodata & \nodata & \nodata &3\\
B012&0.47712&-11.76&14.19&1.851&2.30&\nodata&\nodata&\nodata&\nodata& -1.66 & 6.8 & -1.7 & (14) & 14 &6\\
B020&0.43457&-12.26&13.48&0.827&2.40&\nodata&\nodata&\nodata&\nodata& -0.83 & 6.90 & -0.9 & 8.6 & 14 &3\\
B023&0.55267&-13.66&12.67&0.951&2.40&\nodata&\nodata&\nodata&\nodata& -0.91 & 14.0 & -0.7 & 11.8 & 14 &3\\
B034&0.33445&-11.96&13.28&0.696&2.10&\nodata&\nodata&\nodata&\nodata& -0.59 & 4.4 & -0.6 & (14) & 14 &3\\
B045&0.45788&-11.56&14.30&0.787&2.10&\nodata&\nodata&\nodata&\nodata& -0.76 &10.0& -0.9 & (14) & 14 &3\\
B163&0.47712&-12.86&13.09&0.863&2.50&\nodata&\nodata&\nodata&\nodata& -0.08 & 8.3 & -0.1 & 13.5 & 14 &3\\
B193&0.30750&-12.36&12.74&0.482&2.40&\nodata&\nodata&\nodata&\nodata& -0.2 & 9.6 & -0.1 & 12.9 & 14 &3\\
B218&0.50379&-12.56&13.53&1.121&2.20&\nodata&\nodata&\nodata&\nodata& -0.63 & 8.1 & -0.8 & 8.7 & 14 &3\\
B225&0.32634&-13.46&11.74&0.705&2.30&\nodata&\nodata&\nodata&\nodata& -0.35 & 3.70 & -0.5 & 10.7 & 14 &3\\
B232&0.38561&-11.26&14.24&1.636&2.20&\nodata&\nodata&\nodata&\nodata& -1.49 & 8.9 & -1.9 & (14) & 14 &6\\
B240&0.47567&-11.66&14.29&1.102&2.30&\nodata&\nodata&\nodata&\nodata& -1.74 & 1.79 & -1.5 & (14) & 14 &6\\
B311&0.40140&-11.66&13.91&1.451&1.80&\nodata&\nodata&\nodata&\nodata& -1.58 & 8.5 & \nodata & \nodata & \nodata &6\\
B343&0.43616&-10.46&15.29&1.763&2.40&\nodata&\nodata&\nodata&\nodata& -1.28 & 1.82 & \nodata & \nodata & \nodata &6\\
B373&0.28330&-11.96&13.02&0.598&1.90&\nodata&\nodata&\nodata&\nodata& -0.59 & 11.0 & -0.5 & 13.7 & 14 &3\\
B386&0.35984&-11.56&13.81&0.860&2.20&\nodata&\nodata&\nodata&\nodata& -1.13 & 9.7 & -1.1 & (14) & 14 &6\\
B405&0.58320&-11.86&14.62&0.607&2.50&\nodata&\nodata&\nodata&\nodata& -1.28 & 1.3 & -1.2 & (14) & 14 &6\\
B407&0.38382&-11.36&14.13&0.750&2.00&\nodata&\nodata&\nodata&\nodata& -0.65 & 5.10 & \nodata & \nodata & \nodata &3\\
B158&0.34400&-12.65&12.64&0.970&2.10&\nodata&\nodata&\nodata&\nodata& -0.74 & 2.60 & -0.8 & 9.1 & 14 &3\\
B289&0.50000&-10.50&15.56&1.692&0.20&\nodata&\nodata&\nodata&\nodata& -1.5 & 0.81 & \nodata & \nodata & \nodata &6\\
B338&0.63600&-12.77&13.98&1.734&2.30&\nodata&\nodata&\nodata&\nodata& -0.49 & 8.2 & \nodata & \nodata & \nodata &3\\
B358&0.54900&-11.48&14.83&0.712&1.60&\nodata&\nodata&\nodata&\nodata& -1.85 &10.0& \nodata & \nodata & \nodata &6\\
B379&0.37600&-11.37&14.07&0.533&2.60&\nodata&\nodata&\nodata&\nodata& -0.53 &10.0& -0.4 & 10.5 & 14 &3\\
B384&0.41000&-11.53&14.09&0.770&2.10&\nodata&\nodata&\nodata&\nodata& -0.59 & 10.3 & -0.7 & 13.5 & 14 &3\\
OmegaCen&0.78387&-12.64&14.84&1.631&3.12&0.06&0.66&0.88&3.05& \nodata & \nodata & -1.62 & 11.28 & 9,10&6\\
M54&0.58529&-12.38&14.12&1.041&3.20&0.07&0.70&0.89&3.11& \nodata & \nodata & -1.58 & 10.8 & 9,10&6\\
NGC2808&0.34454&-11.94&13.35&0.796&1.35&0.03&0.70&0.88&3.32& \nodata & \nodata & -1.15 & 8.624 & 9,10&6\\
G1&0.50709&-13.13&12.97&1.640&0.17&0.13&0.57&0.64&3.51& -0.73 & 9 & \nodata & \nodata & 22 &3\\
C11&1.01700&-13.25&15.41&2.383&3.60&0.03&0.66&0.86&4.59& \nodata & \nodata & -1.5 & 14.0 & 22 &6\\
C17&0.87870&-12.97&15.00&2.552&4.90&0.09&0.69&0.87&4.85& \nodata & \nodata & -1.5 & 14.0 & 22 &6\\
C2&0.92985&-12.38&15.84&2.781&0.83&0.11&0.74&0.92&4.29& \nodata & \nodata & -1.4 & 3.50 & 22 &6\\
C21&0.96684&-12.83&15.58&4.311&3.60&0.05&0.66&0.84&4.42& \nodata & \nodata & -1.5 & 14.0 & 22 &6\\
C22&0.70800&-12.39&14.72&2.996&1.30&0.01&0.62&0.78&4.53& \nodata & \nodata & -1.6 & 6.90 & 22 &6\\
C23&0.63837&-14.12&12.64&1.400&6.30&0.10&0.73&0.90&5.94& \nodata & \nodata & -1.5 & 14.0 & 22 &6\\
C25&0.92009&-12.89&15.28&1.254&2.95&0.34&0.88&1.19&5.47& \nodata & \nodata & -0.9 & 14.: & 22 &3\\
C31&0.72380&-12.97&14.22&1.124&4.90&0.10&0.69&0.89&4.82& \nodata & \nodata & -1.5 & 14.0 & 22 &6\\
C32&0.85642&-12.80&15.05&1.984&0.83&0.20&0.74&0.95&4.95& \nodata & \nodata & -1.35 & 3.50 & 22 &6\\
C37&0.63837&-12.65&14.11&1.011&0.95&0.18&0.75&0.92&5.50& \nodata & \nodata & -1.4 & 4.30 & 22 &6\\
C41&0.78179&-12.82&14.66&0.609&4.90&0.11&0.69&0.86&4.70& \nodata & \nodata & -1.5 & 14.0 & 22 &6\\
C7&1.00091&-13.73&14.85&2.349&2.47&0.54&0.91&1.12&6.41& \nodata & \nodata & -0.5 & 14.0 & 22 &3\\
HCH99-18&1.13397&-13.40&15.84&3.879&3.00&-0.05&0.95&1.35&5.99& \nodata & \nodata & -1 & 14.0 & 24 &3\\
HCH99-2&1.05479&-12.58&16.27&3.044&9.00&0.28&0.83&1.02&3.05& \nodata & \nodata & -1.52 & 14.: & 24 &6\\
HGHH92-C11&0.88942&-13.53&14.49&1.607&2.95&0.46&0.93&1.13&4.05& \nodata & \nodata & -0.38 & 14.: & 24 &3\\
HGHH92-C17&0.75376&-13.15&14.19&1.983&1.20&0.20&0.77&0.91&3.26& \nodata & \nodata & -1.35 & 6.00 & 24 &6\\
HGHH92-C21&0.84484&-13.05&14.75&1.908&1.30&0.28&0.78&0.96&3.41& \nodata & \nodata & -1.22 & 6.60 & 24 &6\\
HGHH92-C22&0.57767&-12.39&14.07&1.953&1.30&0.22&0.79&0.94&3.04& \nodata & \nodata & -1.23 & 6.90 & 24 &6\\
HGHH92-C29&0.83294&-13.27&14.47&1.044&1.20&0.47&0.87&1.08&3.94& \nodata & \nodata & -0.59 & 6.00 & 24 &3\\
HGHH92-C36&0.55539&-12.12&14.23&1.834&6.00&0.12&0.74&0.89&2.95& \nodata & \nodata & -1.52 & 14.0 & 24 &6\\
HGHH92-C37&0.45273&-12.65&13.19&0.574&6.40&0.36&0.84&1.02&3.63& \nodata & \nodata & -0.93 & 14.0 & 24 &3\\
HGHH92-C41&0.65685&-13.93&12.93&0.236&1.20&0.46&0.87&1.09&5.04& \nodata & \nodata & -0.58 & 6.00 & 24 &3\\
HGHH92-C7&0.87870&-13.63&14.34&1.626&0.95&0.23&0.75&0.94&3.26& \nodata & \nodata & -1.34 & 4.30 & 24 &6\\
HHH86-C15&0.72380&-12.48&14.71&0.965&1.47&0.42&0.88&1.05&3.60& \nodata & \nodata & -0.7 & 7.80 & 24 &3\\
HHH86-C38&0.45273&-12.33&13.51&0.971&1.30&0.26&0.79&0.95&3.27& \nodata & \nodata & -1.4 & 1.3 & 24 &6\\
R261&0.27664&-12.99&11.97&0.375&1.90&0.38&0.82&1.00&3.69& \nodata & \nodata & -0.95 & 11.0 & \nodata &3\\
W3&1.27261&-15.99&13.94&1.817&0.75&\nodata&0.45&0.64&0.24& \nodata & \nodata & 0 & 0.540 & 15 &3\\
W30&0.97158&-14.32&14.10&1.918&0.45&\nodata&0.41&0.63&0.13& \nodata & \nodata & 0.1 & 0.47 & 15 &3\\
G114&0.64747&-15.56&11.25&0.681&2.40&\nodata&0.46&0.62&3.22& \nodata & \nodata & 0 & 1.10 & 15 &3\\
VUCD3&1.44840&-16.22&14.59&2.982&4.40&0.65&0.94&1.27&4.21& \nodata & \nodata & -0.11 & 13.5 & 16 &3\\
VUCD4&1.47473&-15.18&15.76&2.193&4.50&0.13&0.74&0.99&3.25& \nodata & \nodata & -1.12 & 11 & 16 &3\\
VUCD5&1.35823&-15.62&14.74&1.632&3.40&0.36&0.86&1.11&3.83& \nodata & \nodata & -0.45 & 11 & 16 &3\\
H8005&1.46177&-13.18&17.70&2.628&3.00&\nodata&\nodata&0.96&\nodata& \nodata & \nodata & -1.27 & 13.0 & 17 &6\\
S314&0.54531&-13.74&12.56&2.022&2.90&0.59&0.87&1.10&3.71& \nodata & \nodata & -0.5 & 13.0 & 17 &3\\
S417&1.14737&-14.49&14.82&3.529&5.80&0.20&0.83&1.06&3.52& \nodata & \nodata & -0.7 & 13.0 & 17 &3\\
S490&0.54531&-14.25&12.05&1.839&4.10&\nodata&1.01&1.22&4.26& \nodata & \nodata & 0.18 & 13.0 & 17 &3\\
S928&1.35823&-13.89&16.47&4.601&6.10&0.34&0.99&0.95&3.28& \nodata & \nodata & -1.34 & 13.0 & 17 &6\\
S999&1.30498&-13.37&16.72&8.635&9.40&\nodata&\nodata&0.94&\nodata& \nodata & \nodata & -1.38 & 13.0 & 17 &6\\
UCD2&1.36227&-16.32&14.06&0.603&2.50&\nodata&\nodata&1.12&\nodata& \nodata & \nodata & -0.9 & 10.0 & 22 &3\\
UCD3&1.95555&-16.15&17.19&3.206&4.70&\nodata&\nodata&1.18&\nodata& \nodata & \nodata & -0.52 & 10.0 & 22 &3\\
UCD4&1.48284&-15.28&15.70&2.887&3.40&\nodata&\nodata&1.12&\nodata& \nodata & \nodata & -0.85 & 10.0 & 22 &3\\
UCD5&1.48284&-14.82&16.16&2.718&10.98&\nodata&\nodata&1.00&\nodata& \nodata & \nodata & -0.6: & 14.0 & 22 &3\\
F-5&0.66330&-14.54&12.34&1.588&3.80&\nodata&\nodata&1.33&\nodata& \nodata & \nodata & -0.34 & 15.0 & 21 &3\\
F-6&0.86742&-13.94&13.96&2.651&1.60&\nodata&\nodata&0.72&\nodata& \nodata & \nodata & -1.31 & 11.0 & 21 &6\\
F-7&1.16845&-13.94&15.47&2.711&2.40&\nodata&\nodata&0.88&\nodata& \nodata & \nodata & -1.2 & 14.8 & 21 &6\\
F-9&0.96433&-14.14&14.25&2.362&3.20&\nodata&\nodata&1.20&\nodata& \nodata & \nodata & -0.62 & 15.0 & 21 &3\\
F-11&0.56639&-14.34&12.06&0.842&3.20&\nodata&\nodata&1.10&\nodata& \nodata & \nodata & -0.61 & 15.0 & 21 &3\\
F-12&1.00572&-14.34&14.25&1.998&2.90&\nodata&\nodata&1.13&\nodata& \nodata & \nodata & -0.4 & 13.0 & 21 &3\\
F-17&0.56639&-14.14&12.26&1.196&3.30&\nodata&\nodata&1.09&\nodata& \nodata & \nodata & -0.55 & 15.0 & 21 &3\\
F-22&1.00572&-13.94&14.65&3.286&3.40&\nodata&\nodata&1.19&\nodata& \nodata & \nodata & -0.49 & 15.0 & 21 &3\\
F-34&0.56639&-13.54&12.86&1.498&2.90&\nodata&\nodata&1.11&\nodata& \nodata & \nodata & -0.77 & 14.9 & 21 &3\\
F-51&0.66330&-13.44&13.44&1.710&4.20&\nodata&\nodata&1.16&\nodata& \nodata & \nodata & -0.23 & 15.0 & 21 &3\\
F-53&0.66330&-13.44&13.44&1.422&2.90&\nodata&\nodata&1.16&\nodata& \nodata & \nodata & -0.8 & 13.8 & 21 &3\\
F-59&0.74248&-13.34&13.94&0.438&10.98&\nodata&\nodata&1.02&\nodata& \nodata & \nodata & -0.7: & 14.: & 21 &3\\
M59cO&1.49956&-15.15&15.92&9.870&5.90&\nodata&\nodata&1.08&\nodata& \nodata & \nodata & -0.03 & 9.30 & 20 &3\\
VCC1073Nucleus&0.98465&-14.16&14.33&2.521&10.98&\nodata&\nodata&1.07& \nodata & \nodata &\nodata&  & \nodata & \nodata &3\\
VCC1254Nucleus&1.11934&-15.57&13.60&2.205&10.98&\nodata&\nodata&1.01&\nodata& \nodata & \nodata & -1.5 & 14.: & 19 &6\\
LeoI&2.51880&-14.17&21.99&6.555&4.60&0.15&0.80&\nodata&3.07& \nodata & \nodata & -1.4 & 13: & 18 &6\\
LeoII&2.26636&-11.75&23.15&19.803&17.00&\nodata&0.65&\nodata&2.80& \nodata & \nodata & -1.6 & 13: & 18 &4\\
UrsaMinor&2.47718&-10.86&25.09&140.587&79.00&-0.10&1.30&\nodata&3.26& \nodata & \nodata & -1.9 & 13: & 18 &2\\
Sculptor&2.20377&-13.31&21.28&3.966&3.00&\nodata&0.70&\nodata&2.91& \nodata & \nodata & -1.5 & 13: & 18 &6\\
Carina&2.46228&-11.32&24.56&47.583&31.00&\nodata&0.70&\nodata&2.72& \nodata & \nodata & -1.8 & 13: & 18 &4\\
Draco&2.36135&-10.70&24.67&130.304&84.00&0.10&0.95&\nodata&2.85& \nodata & \nodata & -2 & 13: & 18 &2\\
Sextans&2.79944&-11.47&26.10&85.219&39.00&\nodata&0.70&\nodata&2.67& \nodata & \nodata & -1.9 & 13: & 18 &2\\
Fornax&2.60203&-13.38&23.20&23.473&34.40&0.08&0.63&1.02&0.81& \nodata & \nodata & -1.2 & 13: & 18 &4\\
M32&1.99264&-18.44&15.09&2.859&5.60&0.64&0.99&\nodata&2.73& \nodata & \nodata & -1.1 & 13: & 23 &5\\
NGC1052&3.51325&-23.79&17.34&2.513&1.28&0.41&0.90&1.01&3.88& -0.1 & 5.1 & 0.08 & 9.7 & 5 &1\\
NGC1395&3.87145&-24.81&18.11&3.525&2.71&0.56&0.92&1.03&3.59& 0.1 & 5.6 & 0.43 & 3.8 & 3 &1\\
NGC1439&3.45960&-23.08&17.79&2.467&1.11&0.37&0.84&0.94&3.58& -0.2 & 4.0: & -0.15 &10.0& 6 &1\\
NGC1549&3.66308&-24.31&17.58&2.193&2.10&0.49&0.91&1.01&3.83& 0 & 5.4 & 0.37 & 3.1 & 3 &1\\
NGC3193&3.55733&-23.88&17.48&2.546&1.52&0.44&0.91&1.02&3.89& -0.1 & 5.4 & 0.18 & 5.1 & 3 &1\\
NGC3377&3.34707&-23.26&17.04&1.128&0.76&0.28&0.82&0.92&3.55& -0.1 & 3.7: & 0.10 &10.0& 3,4 &1\\
NGC3379&3.85417&-24.91&17.93&1.879&2.26&0.51&0.93&1.04&3.85& 0.1 & 5.8: & 0.10 &10.0& 3,4 &1\\
NGC4125&3.84357&-24.66&18.13&2.998&1.93&0.48&0.90&1.01&3.68& 0.1 & 5.1 & \nodata & \nodata & \nodata &1\\
NGC4168&3.72405&-24.27&17.91&2.040&1.57&0.45&0.87&0.98&3.49& 0 & 4.6 & \nodata & \nodata & \nodata &1\\
NGC4278&3.41698&-23.36&17.30&5.033&1.37&0.42&0.89&1.00&3.74& -0.2 & 4.9 & 0.27 & 2.9 & 3,4 &1\\
NGC4473&4.16626&-25.57&18.82&1.636&2.98&0.59&0.91&1.02&3.76& 0.2 & 5.4 & 0.30 & 2.8 & 3 &1\\
NGC4478&3.38480&-23.45&17.04&1.351&3.13&0.60&0.87&0.98&3.68& -0.1 & 4.6: & \nodata & 9.9 & \nodata &1\\
NGC4494&3.88502&-24.75&18.24&0.895&1.49&0.44&0.85&0.95&3.58& 0.1 & 4.2 & 0.26 & 2.6 & 3 &1\\
NGC4636&3.90000&-24.80&18.27&2.095&1.31&0.42&0.90&1.01&3.88& 0.1 & 5.1: & 0.06 & 13.5 & 4,5 &1\\
NGC4742&3.38187&-23.32&17.16&0.589&0.69&0.27&0.76&0.85&3.54& -0.1 & 2.9: & -0.01 &10.0& 4 &1\\
NGC4889&4.73030&-26.43&20.79&12.562&2.56&0.55&0.97&1.09&3.95& 0.3 & 4.8: & 0.30 &10.0& 4 &1\\
NGC5576&3.60455&-24.05&17.54&2.019&1.13&0.38&0.85&0.95&3.81& -0.1 & 4.2 & 0.31 & 2.2 & 3 &1\\
NGC5831&3.51436&-23.58&17.56&2.000&2.45&0.53&0.90&1.00&3.73& -0.1 & 5.1 & -0.10 & 8.8 & 4,5 &1\\
NGC584&3.68645&-24.27&17.73&2.689&1.69&0.46&0.90&1.01&3.83& 0 & 5.1 & 0.20 & 10.0 & 3,4 &1\\
NGC636&3.46846&-23.23&17.68&2.188&1.78&0.47&0.91&1.01&3.78& -0.2 & 5.4 & -0.07 & 11.0 & 4,6 &1\\
NGC6411&3.72629&-24.42&17.78&1.315&1.25&0.41&0.85&0.95&3.41& 0.1 & 4.2: & 0.20 &10.0& 4 &1\\
NGC6909&3.49353&-23.56&17.48&1.559&0.51&0.24&0.81&0.90&3.24& -0.1 & 3.6 & \nodata & \nodata & \nodata &1\\
NGC720&3.64221&-24.27&17.51&3.134&1.74&0.46&0.95&1.06&3.79& 0 & 6.3 & 0.04 & 7.8 & 3 &1\\
NGC7507&3.56552&-23.90&17.50&3.443&2.34&0.52&0.92&1.03&3.87& -0.1 & 5.6 & 0.25 & 3.9 & 3 &1\\
NGC7619&4.08524&-25.31&18.69&6.241&2.22&0.50&0.93&1.04&3.69& 0.2 & 5.8: & 0.27 & 13.2 & 6 &1\\
NGC7626&4.01909&-25.07&18.60&3.221&2.29&0.51&0.95&1.06&3.81& 0.1 & 6.3: & -0.00 & 18 & 6 &1\\
NGC7785&3.92840&-24.92&18.29&4.627&2.18&0.50&0.95&1.06&3.91& 0.1 & 6.3 & \nodata & \nodata & \nodata &1\\
IC794&3.03621&-20.53&18.22&1.081&1.52&\nodata&\nodata&\nodata&3.24& -0.5 & 5.3: & -0.45 &10.0& 4 &5\\
NGC4515&2.99951&-21.31&17.25&1.447&1.53&\nodata&\nodata&\nodata&3.25& -0.4 & 5.3 & \nodata & \nodata & \nodata &5\\
VCC351&2.85191&-19.71&18.12&2.353&2.22&\nodata&\nodata&\nodata&3.36& -0.7 & 6.5 & \nodata & 3.36 & \nodata &5\\
VCC1087&2.98000&-20.24&18.23&1.251&1.39&\nodata&\nodata&\nodata&3.21& -0.7 & 5 & -0.27 & 5.3 & 7,8 &5\\
VCC1122&2.94900&-19.54&18.77&2.159&0.75&\nodata&\nodata&\nodata&3.02& -0.8 & 4 & -0.61 & 2.5 & 7 &5\\
VCC1254&3.09100&-19.59&19.43&1.739&1.72&\nodata&\nodata&\nodata&3.30& -0.7 & 5.6 & -0.41 & 6.6 & 8 &5\\
VCC1261&3.05000&-20.40&18.42&0.481&0.43&\nodata&\nodata&\nodata&2.68& -0.5 & 3.7 & -0.27 & 2.5 & 1,7 &5\\
VCC1407&2.98800&-18.88&19.63&3.833&0.36&0.30&\nodata&\nodata&2.48& -0.7 & 3.6 & -0.70 & 5 & 7,8 &5\\
VCC1491&2.99300&-19.35&19.18&2.380&0.68&0.30&\nodata&\nodata&2.97& -0.7 & 3.9 & -0.44 & 5.6 & 7,8 &5\\
VCC543&3.06800&-19.36&19.55&2.889&0.45&\nodata&\nodata&\nodata&2.72& -0.6 & 5.6 & -0.49 & 5.4 & 7 &5\\
VCC856&3.13000&-19.99&19.23&1.052&0.81&\nodata&\nodata&\nodata&3.06& -0.6 & 4.1 & -0.43 & 5 & 7 &5\\
VCC929&3.14600&-21.10&18.20&0.586&0.73&\nodata&\nodata&\nodata&3.01& -0.4 & 4 & -0.50 & 4.3 & 1,8 &5\\
E208-G021&3.36447&-23.21&17.18&2.060&1.07&0.35&0.81&0.91&3.60& -0.2 & 3.6 & \nodata &  & \nodata &1\\
E221-G026&3.64360&-24.20&17.59&1.054&1.12&0.37&0.85&0.95&3.65& 0 & 4.2 & \nodata &  & \nodata &1\\
E318-G021&3.85991&-24.71&18.16&1.349&2.18&0.50&0.91&1.02&3.71& 0.1 & 5.4 & \nodata &  & \nodata &1\\
E462-G015&4.22745&-25.58&19.12&5.071&1.07&0.36&0.91&1.01&3.78& 0.2 & 5.4 & \nodata &  & \nodata &1\\
E467-G054&3.88710&-24.78&18.22&3.611&2.47&0.53&0.94&1.06&3.98& 0 & 6 & \nodata &  & \nodata &1\\
E507-G025&4.05711&-25.13&18.72&3.574&1.01&0.33&0.88&0.98&3.90& 0.1 & 4.7 & \nodata &  & \nodata &1\\
IC1459&3.82304&-24.80&17.88&4.525&2.19&0.50&0.95&1.06&4.06& 0 & 6.3 & 0.33 & 8 & 5 &1\\
IC2311&3.55336&-23.94&17.40&2.801&1.06&0.35&0.85&0.96&3.53& 0 & 4.2 & \nodata &  & \nodata &1\\
IC310&4.09244&-25.24&18.79&3.185&1.89&0.47&0.94&1.05&3.98& 0.1 & 6 & \nodata &  & \nodata &1\\
IC4051&3.88442&-24.81&18.18&2.713&1.80&0.47&0.93&1.04&3.45& 0.1 & 5.8 & 0.13 &10.0& 4 &1\\
IC4296&4.45701&-26.20&19.65&5.931&2.31&0.51&0.91&1.02&3.77& 0.3 & 3.8 & 0.35 & 5.2 & 5 &1\\
IC4889&3.85561&-24.54&18.31&1.807&1.22&0.40&0.87&0.98&3.70& 0 & 4.6 & 0.18 & 5.5 & 6 &1\\
IC5328&3.84270&-24.72&18.06&2.089&1.36&0.42&0.92&1.03&3.88& 0 & 5.6 & \nodata & \nodata & \nodata &1\\
NGC0016&3.57064&-24.11&17.31&1.680&1.06&0.35&0.90&1.01&3.98& -0.1 & 5.1 & \nodata & \nodata & \nodata &1\\
NGC0315&4.46523&-26.05&19.84&8.179&2.66&0.56&0.93&1.04&3.92& 0.25 & 8.2 & \nodata & \nodata & \nodata &1\\
NGC0474&3.54290&-23.64&17.65&2.145&1.07&0.36&0.80&0.90&3.65& -0.1 & 2.5 & \nodata & \nodata & \nodata &1\\
NGC0584&3.68645&-24.27&17.73&2.689&1.69&0.46&0.90&1.01&3.83& -0.03 & 5.5 & \nodata & \nodata & \nodata &1\\
NGC0636&3.46846&-23.23&17.68&2.180&1.78&0.47&0.91&1.01&3.78& -0.19 & 5.8 & \nodata & \nodata & \nodata &1\\
NGC0777&4.25774&-25.70&19.16&6.895&2.82&0.58&0.95&1.06&3.87& 0.2 & 4.5 & 0.36 & 5.4 & 5 &1\\
NGC0936&3.67039&-24.15&17.77&2.284&2.28&0.51&0.90&1.01&4.12& -0.09 & 6.6 & \nodata & \nodata & \nodata &1\\
NGC1016&4.47692&-26.14&19.81&4.208&2.70&0.56&0.92&1.03&3.85& 0.3 & 4 & \nodata & \nodata & \nodata &1\\
NGC1060&4.39185&-25.97&19.56&6.542&2.39&0.52&0.95&1.06&3.95& 0.2 & 4.5 & \nodata & \nodata & \nodata &1\\
NGC1175&3.96888&-25.09&18.33&2.061&1.48&0.44&0.81&0.91&4.08& 0.1 & 3.6 & \nodata & \nodata & \nodata &1\\
NGC1199&3.11715&-24.02&15.13&0.825&1.25&0.41&0.94&1.05&3.57& 0 & 6.0: & -0.06 & 11 & 2 &1\\
NGC1209&3.68751&-24.29&17.72&2.494&2.03&0.49&0.90&1.01&3.84& 0 & 5.1 & 0.41 & 4.8 & 5 &1\\
NGC1270&3.78878&-24.65&17.86&7.375&3.01&0.60&0.99&1.10&4.05& 0 & 7.4 & \nodata & \nodata & \nodata &1\\
NGC1272&3.85692&-24.81&18.04&4.556&1.81&0.47&0.89&0.99&3.56& 0.1 & 4.9 & \nodata & \nodata & \nodata &1\\
NGC1283&3.78453&-24.58&17.91&2.976&2.53&0.54&0.93&1.04&3.80& 0 & 5.8 & \nodata & \nodata & \nodata &1\\
NGC1344&3.42878&-23.39&17.32&1.810&1.41&0.43&0.85&0.95&3.73& -0.2 & 4.2 & \nodata & \nodata & \nodata &1\\
NGC1404&3.96895&-25.20&18.21&2.346&2.67&0.56&0.94&1.05&4.00& 0.1 & 6 & 0.27 & 5.8 & 3 &1\\
NGC1453&4.17557&-25.43&19.01&5.048&2.38&0.52&0.91&1.02&3.99& 0.1 & 5.4 & 0.24 & 9.4 & 5 &1\\
NGC1521&4.00509&-25.06&18.53&2.847&1.73&0.46&0.89&1.00&3.53& 0.2 & 4.9 & 0.27 & 3.2 & 5 &1\\
NGC1537&3.43373&-23.53&17.20&1.582&1.22&0.40&0.85&0.95&3.63& -0.1 & 4.2 & 0.35 & 2.8 & \nodata &1\\
NGC1553&3.70125&-24.54&17.53&1.322&1.72&0.46&0.84&0.94&3.93& 0 & 4 & 0.19 & 4.8 & 5 &1\\
NGC1587&3.99523&-24.97&18.57&3.031&1.31&0.41&0.92&1.02&3.87& 0.1 & 5.6 & \nodata & \nodata & \nodata &1\\
NGC1595&3.70940&-24.28&17.83&0.809&1.66&0.45&0.90&1.01&3.83& 0 & 5.1 & \nodata & \nodata & \nodata &1\\
NGC1601&3.49748&-23.28&17.77&1.276&1.52&0.44&0.89&1.00&3.71& -0.2 & 4.9 & \nodata & \nodata & \nodata &1\\
NGC1653&3.92619&-24.87&18.33&2.744&1.76&0.46&0.86&0.96&3.71& 0.1 & 4.4 & \nodata & \nodata & \nodata &1\\
NGC1726&3.98752&-25.07&18.44&2.444&1.68&0.46&0.86&0.96&3.73& 0.1 & 4.4 & \nodata & \nodata & \nodata &1\\
NGC1930&3.81999&-24.59&18.07&2.174&1.30&0.41&0.91&1.02&3.88& 0 & 5.4 & \nodata & \nodata & \nodata &1\\
NGC2314&3.79838&-24.72&17.84&4.116&2.33&0.52&0.91&1.02&4.02& 0 & 5.4 & \nodata & \nodata & \nodata &1\\
NGC2329&4.10445&-25.13&18.96&4.948&1.85&0.47&0.90&1.01&3.77& 0.1 & 5.1: & 0.30 &10.0& 4 &1\\
NGC2434&3.51563&-23.69&17.46&2.748&0.69&0.27&0.82&0.92&3.43& -0.1 & 3.7 & 0.40 & 5.6 & 6 &1\\
NGC2634&3.50396&-23.23&17.86&3.194&1.70&0.46&0.88&0.99&3.52& -0.1 & 4.7 & \nodata & \nodata & \nodata &1\\
NGC2639&3.95907&-25.00&18.37&1.767&0.87&0.30&0.84&0.94&3.96& 0.1 & 4 & \nodata & \nodata & \nodata &1\\
NGC2672&4.28794&-25.65&19.35&5.007&3.36&0.62&0.95&1.07&4.15& 0.1 & 6.3 & \nodata & \nodata & \nodata &1\\
NGC2693&4.25674&-25.62&19.23&4.747&3.32&0.62&0.89&1.00&4.07& 0.2 & 4.9: & \nodata &10.0& 4 &1\\
NGC2768&3.78575&-24.56&17.94&2.146&1.35&0.42&0.91&1.02&3.62& 0.1 & 5.4 & 0.19 & 3.3 & 3 &1\\
NGC2865&3.83118&-24.57&18.15&1.704&1.04&0.35&0.80&0.90&3.61& 0.1 & 3.4 & \nodata & \nodata & \nodata &1\\
NGC2872&3.84854&-24.68&18.13&4.284&2.18&0.50&0.93&1.04&3.93& 0 & 5.8 & \nodata & \nodata & \nodata &1\\
NGC2880&3.40663&-23.18&17.42&1.684&1.29&0.41&0.86&0.96&3.69& -0.2 & 4.4 & \nodata & \nodata & \nodata &1\\
NGC2904&3.55090&-23.64&17.69&4.011&1.37&0.42&0.91&1.01&3.75& -0.1 & 5.4 & 0.31 & 7.8 & 6 &1\\
NGC2974&4.27865&-26.34&18.62&1.631&2.40&0.52&0.93&1.04&5.37& 0.1 & 5.8: & 0.02 & 13.9 & 5 &1\\
NGC2986&3.98934&-25.15&18.36&4.046&2.40&0.52&0.89&1.00&3.78& 0.1 & 4.9 & 0.41 & 3.3 & 3 &1\\
NGC3070&4.01873&-25.20&18.46&2.426&1.22&0.40&0.87&0.98&3.77& 0.1 & 4.6 & \nodata & \nodata & \nodata &1\\
NGC3078&3.97638&-25.06&18.39&3.032&2.45&0.53&0.92&1.03&3.86& 0.1 & 5.6 & \nodata & \nodata & \nodata &1\\
NGC3087&3.95908&-24.93&18.43&4.332&1.76&0.46&0.92&1.03&4.03& 0.1 & 5.6 & \nodata & \nodata & \nodata &1\\
NGC3115&3.77932&-24.85&17.61&1.995&1.90&0.48&0.90&1.01&4.01& 0 & 5.1 & 0.40 & 3.0 & 3 &1\\
NGC3136&3.95828&-24.95&18.41&3.123&0.92&0.31&0.76&0.85&3.59& 0.1 & 2.9 & 0.66 & 1.5 & 5 &1\\
NGC3158&4.58806&-26.18&20.32&7.380&2.88&0.58&0.93&1.04&3.94& 0.3 & 4.1 & \nodata & \nodata & \nodata &1\\
NGC323&4.24532&-25.64&19.15&6.849&2.60&0.55&0.95&1.06&3.94& 0.2 & 6.3 & \nodata & \nodata & \nodata &1\\
NGC3250&4.13283&-25.45&18.78&3.725&2.64&0.56&0.92&1.03&3.99& 0.1 & 5.6 & 0.41 & 2.8 & 3 &1\\
NGC3257&3.44776&-23.28&17.53&1.279&1.32&0.42&0.87&0.98&3.62& -0.2 & 4.6 & \nodata & \nodata & \nodata &1\\
NGC3258&3.89433&-24.84&18.20&4.040&1.87&0.47&0.90&1.01&3.83& 0.1 & 5.1 & 0.38 & 4.5 & 5 &1\\
NGC3268&4.03052&-25.01&18.72&3.320&1.70&0.46&0.92&1.03&3.72& 0.1 & 5.6 & 0.61 & 9.8 & 5 &1\\
NGC3305&3.69554&-24.37&17.68&2.541&2.36&0.52&0.91&1.01&3.91& 0 & 5.4 & -0.20 & 1.2 & 2 &1\\
NGC3308&3.82038&-24.66&18.01&1.951&2.33&0.52&0.91&1.02&4.05& 0 & 5.4 & \nodata & \nodata & \nodata &1\\
NGC3309&4.12307&-25.45&18.73&3.285&2.67&0.56&0.89&1.00&3.72& 0.2 & 4.9 & \nodata & \nodata & \nodata &1\\
NGC3311&4.28876&-25.56&19.45&2.589&2.25&0.51&0.88&0.99&4.33& 0.1 & 4.7 & \nodata & \nodata & \nodata &1\\
NGC3348&3.98987&-25.03&18.49&3.230&1.26&0.41&0.92&1.03&3.63& 0.1 & 5.6 & \nodata & \nodata & \nodata &1\\
NGC3557&4.47813&-26.12&19.83&5.441&2.40&0.52&0.90&1.01&3.76& 0.3 & 3.7 & 0.24 & 5.8 & 5 &1\\
NGC3562&4.13012&-25.47&18.75&4.182&2.40&0.52&0.87&0.97&3.53& 0.2 & 3.3 & \nodata & \nodata & \nodata &1\\
NGC3585&3.94132&-25.19&18.08&2.125&1.76&0.46&0.89&1.00&3.84& 0.1 & 4.9 & \nodata & \nodata & \nodata &1\\
NGC3607&3.64861&-24.26&17.55&3.230&1.74&0.46&0.90&1.01&3.85& 0 & 5.1 & 0.38 & 3.1 & 5 &1\\
NGC3608&3.51139&-23.59&17.53&2.960&1.12&0.38&0.91&1.02&3.37& -0.1 & 5.4: & 0.16 &10.0& 3,4 &1\\
NGC3610&3.63939&-24.11&17.65&1.486&1.71&0.46&0.83&0.93&3.65& 0 & 3.9 & 0.40 & 1.6 & 3 &1\\
NGC3640&3.78960&-24.20&18.32&2.390&2.11&0.50&0.87&0.97&3.62& 0 & 4.6 & -0.01 & 8.3 & 6 &1\\
NGC3706&4.13109&-25.38&18.85&4.528&1.98&0.48&0.92&1.02&4.03& 0.1 & 5.6 & \nodata & \nodata & \nodata &1\\
NGC380&3.77027&-24.50&17.91&4.280&2.40&0.52&0.94&1.06&3.92& 0 & 6 & \nodata & \nodata & \nodata &1\\
NGC3837&3.88307&-24.77&18.21&3.734&2.68&0.56&0.92&1.03&4.12& 0 & 5.6 & \nodata & \nodata & \nodata &1\\
NGC3842&4.29826&-25.72&19.34&4.910&2.19&0.50&0.88&0.98&3.54& 0.2 & 3.4 & \nodata & \nodata & \nodata &1\\
NGC385&3.67339&-24.11&17.82&2.063&1.11&0.37&0.86&0.96&3.66& 0 & 4.4 & \nodata & \nodata & \nodata &1\\
NGC3862&4.17040&-25.37&19.05&4.321&2.36&0.52&0.92&1.03&4.04& 0.1 & 5.6 & \nodata & \nodata & \nodata &1\\
NGC3873&3.90985&-24.74&18.38&3.657&1.29&0.41&0.92&1.03&3.93& 0 & 5.6 & \nodata & \nodata & \nodata &1\\
NGC3904&3.72108&-24.41&17.76&2.510&1.79&0.47&0.89&1.00&3.80& 0 & 4.9 & \nodata & \nodata & \nodata &1\\
NGC392&3.68896&-24.34&17.67&3.655&1.60&0.45&0.86&0.96&3.83& 0 & 4.4 & \nodata & \nodata & \nodata &1\\
NGC3923&4.26203&-25.77&19.10&2.509&2.49&0.54&0.91&1.01&3.92& 0.2 & 3.8 & \nodata & \nodata & \nodata &1\\
NGC3962&3.81827&-24.70&17.96&2.306&1.78&0.47&0.89&1.00&3.73& 0.1 & 4.9: & 0.08 &10.0& 5 &1\\
NGC4036&3.58517&-24.11&17.38&1.983&2.27&0.51&0.85&0.95&3.86& -0.1 & 4.2 & \nodata & \nodata & \nodata &1\\
NGC410&4.32571&-25.80&19.40&6.278&2.41&0.53&0.93&1.04&3.83& 0.2 & 4.1 & \nodata & \nodata & \nodata &1\\
NGC4105&3.93990&-24.94&18.32&3.481&1.50&0.44&0.87&0.98&3.77& 0.1 & 4.6 & \nodata & \nodata & \nodata &1\\
NGC4125&3.84357&-24.66&18.13&2.998&1.93&0.48&0.90&1.01&3.68& 0.1 & 5.1 & 0.18 & 3.5 & 3 &1\\
NGC4169&3.76460&-24.65&17.74&2.359&1.34&0.42&0.87&0.97&3.94& 0 & 4.6 & \nodata & \nodata & \nodata &1\\
NGC4261&4.13970&-25.48&18.78&4.563&2.30&0.51&0.95&1.07&3.99& 0.1 & 6.3 & 0.40 &10.0& 3,4 &1\\
NGC4365&3.95596&-25.04&18.31&3.192&1.99&0.48&0.93&1.04&3.76& 0.1 & 5.8 & 0.30 & 3.8 & 3,4 &1\\
NGC4374&3.97980&-25.19&18.28&3.973&5.06&0.66&0.93&1.04&3.69& 0.1 & 5.8 & 0.10 & 9.8 & 4,5 &1\\
NGC4472&4.33547&-25.91&19.33&4.606&2.46&0.53&0.93&1.04&3.77& 0.2 & 4.1 & 0.41 & 3.4 & 3,4 &1\\
NGC4594&4.49484&-26.40&19.64&2.492&1.46&0.43&0.88&0.98&3.77& 0.3 & 3.4: & 0.20 &10.0& 4 &1\\
NGC4621&3.40263&-23.27&17.31&4.299&1.61&0.45&0.90&1.01&3.78& -0.2 & 5.1 & 0.40 & 3.1 & 3,4 &1\\
NGC4660&3.35340&-23.22&17.11&2.723&1.18&0.39&0.88&0.98&3.70& -0.2 & 4.7 & 0.34 & 3.8 & 3 &1\\
NGC4697&4.00507&-25.30&18.29&1.251&1.10&0.37&0.87&0.97&3.74& 0.2 & 4.6: & 0.20 &10.0& 4,5 &1\\
NGC4816&3.99281&-25.25&18.28&3.007&1.76&0.46&0.87&0.98&3.79& 0.1 & 4.6 & \nodata & \nodata & \nodata &1\\
NGC4839&4.34453&-25.90&19.39&3.881&2.75&0.57&0.88&0.99&3.72& 0.2 & 3.4: & 0.30 & 10.0 & 4 &1\\
NGC4864&3.64160&-24.34&17.43&1.889&2.89&0.58&0.94&1.06&3.78& 0 & 6 & -0.03 &10.0& 4 &1\\
NGC4874&4.63978&-26.20&20.57&5.210&2.44&0.53&0.87&0.98&3.78& 0.3 & 3.3: & 0.30 &10.0& 4 &1\\
NGC4881&3.82521&-24.52&18.17&2.976&2.32&0.51&0.94&1.06&3.97& 0 & 6 & \nodata & \nodata & \nodata &1\\
NGC4923&3.60511&-23.76&17.83&2.775&1.94&0.48&0.89&0.99&3.86& -0.1 & 4.9 & \nodata & \nodata & \nodata &1\\
NGC4976&3.89019&-24.98&18.04&1.128&0.72&0.28&0.81&0.91&3.34& 0.2 & 3.6 & \nodata & \nodata & \nodata &1\\
NGC499&3.95729&-25.02&18.34&2.990&2.91&0.59&0.93&1.04&4.23& 0 & 5.8 & \nodata & \nodata & \nodata &1\\
NGC5018&4.06582&-25.43&18.47&1.996&1.25&0.41&0.80&0.90&3.54& 0.2 & 2.6 & 0.01 & 3 & 6 &1\\
NGC5061&4.00662&-25.26&18.34&1.876&1.12&0.38&0.83&0.93&3.61& 0.2 & 3.9 & \nodata & \nodata & \nodata &1\\
NGC507&4.29650&-25.72&19.33&8.144&1.91&0.48&0.89&0.99&3.95& 0.2 & 4.9: & 0.20 &10.0& 4 &1\\
NGC5077&3.95652&-24.94&18.41&4.318&2.47&0.53&0.95&1.07&3.87& 0.1 & 6.3: & 0.08 & 15 & 5 &1\\
NGC5087&3.77656&-24.50&17.95&4.682&1.37&0.42&0.89&1.00&3.97& 0 & 4.9 & \nodata & \nodata & \nodata &1\\
NGC5129&4.35169&-25.71&19.61&5.296&1.34&0.42&0.83&0.93&3.57& 0.2 & 2.9 & \nodata & \nodata & \nodata &1\\
NGC5322&3.89415&-24.86&18.17&2.664&1.79&0.47&0.88&0.98&3.81& 0.1 & 4.7 & 0.40 & 1.9 & 3 &1\\
NGC533&4.39377&-25.84&19.70&6.962&2.65&0.56&0.94&1.05&3.85& 0.2 & 4.3 & \nodata & \nodata & \nodata &1\\
NGC5380&3.55933&-23.97&17.39&1.401&1.17&0.39&0.91&1.01&3.85& -0.1 & 5.4 & \nodata & \nodata & \nodata &1\\
NGC5485&3.59959&-23.91&17.66&1.558&1.96&0.48&0.85&0.95&3.91& -0.1 & 4.2 & \nodata & \nodata & \nodata &1\\
NGC5490&4.10988&-25.30&18.82&5.289&2.58&0.55&0.94&1.05&3.97& 0.1 & 6 & \nodata & \nodata & \nodata &1\\
NGC5557&3.95036&-25.25&18.06&2.858&2.29&0.51&0.87&0.98&3.78& 0.1 & 4.6 & \nodata & \nodata & \nodata &1\\
NGC5576&3.60455&-24.05&17.54&2.019&1.13&0.38&0.85&0.95&3.81& -0.1 & 4.2 & 0.31 & 2.2 & 3 &1\\
NGC5761&3.79494&-24.57&17.98&1.867&1.00&0.33&0.80&0.89&3.62& 0.1 & 3.4 & \nodata & \nodata & \nodata &1\\
NGC5791&3.96572&-25.01&18.39&2.643&1.79&0.47&0.89&1.00&3.86& 0.1 & 4.9 & \nodata & \nodata & \nodata &1\\
NGC5796&3.95158&-24.97&18.36&3.455&2.41&0.53&0.94&1.05&4.12& 0 & 6 & 0.12 &10.0& 4 &1\\
NGC5813&3.95979&-24.95&18.41&3.219&2.19&0.50&0.92&1.03&3.85& 0.1 & 5.6: & -0.05 & 11.7 & 4,5 &1\\
NGC5898&3.72592&-24.42&17.78&2.504&1.79&0.47&0.90&1.01&3.76& 0 & 5.1 & 0.18 & 7.7 & 5 &1\\
NGC596&3.54750&-23.70&17.61&1.593&1.19&0.39&0.85&0.95&3.70& -0.1 & 4.2 & 0.16 & 4.4 & 3 &1\\
NGC5982&3.94717&-24.96&18.35&3.907&2.04&0.49&0.87&0.97&3.72& 0.1 & 4.6 & \nodata & \nodata & \nodata &1\\
NGC6020&3.62890&-24.14&17.57&2.195&0.25&0.14&0.93&1.04&3.67& 0 & 5.8 & \nodata & \nodata & \nodata &1\\
NGC6086&4.25167&-25.63&19.20&5.555&1.85&0.47&0.89&1.00&3.55& 0.2 & 3.6 & \nodata & \nodata & \nodata &1\\
NGC6107&4.56710&-26.13&20.27&3.103&2.32&0.52&0.98&1.10&5.26& 0 & 7.1 & \nodata & \nodata & \nodata &1\\
NGC6109&4.07277&-25.10&18.83&4.020&2.27&0.51&0.95&1.07&3.41& 0.2 & 6.3 & \nodata & \nodata & \nodata &1\\
NGC6146&4.40158&-25.93&19.65&4.767&2.35&0.52&0.90&1.01&3.86& 0.2 & 3.7 & \nodata & \nodata & \nodata &1\\
NGC6158&3.97850&-24.88&18.58&2.232&1.65&0.45&0.89&1.00&3.96& 0.1 & 4.9 & \nodata & \nodata & \nodata &1\\
NGC6173&4.48185&-26.05&19.93&4.725&2.38&0.52&0.92&1.03&3.63& 0.3 & 4 & \nodata & \nodata & \nodata &1\\
NGC661&3.65138&-24.17&17.65&1.338&1.53&0.44&0.87&0.97&3.59& 0 & 4.6 & \nodata & \nodata & \nodata &1\\
NGC6702&3.96212&-24.81&18.56&2.148&1.43&0.43&0.84&0.94&3.56& 0.1 & 4.0: & 0.12 &10.0& 4 &1\\
NGC6703&3.70461&-24.26&17.83&1.751&1.79&0.47&0.89&1.00&3.76& 0 & 4.9 & 0.10 &10.0& 4 &1\\
NGC6721&3.98157&-25.00&18.48&4.059&1.27&0.41&0.94&1.05&3.89& 0.1 & 6 & 0.31 & 5 & 5 &1\\
NGC6758&3.93604&-24.95&18.30&3.726&1.41&0.43&0.94&1.06&3.94& 0.1 & 6.0: & -0.01 & 16 & 5 &1\\
NGC6776&4.10080&-25.32&18.75&2.307&1.49&0.44&0.86&0.96&3.71& 0.2 & 4.4 & 0.22 & 2.7 & 5 &1\\
NGC679&3.91359&-24.94&18.19&2.774&2.43&0.53&0.89&0.99&4.10& 0 & 4.9 & \nodata & \nodata & \nodata &1\\
NGC680&3.30190&-23.96&16.12&1.274&1.39&0.43&0.90&1.00&3.82& -0.1 & 5.1 & \nodata & \nodata & \nodata &1\\
NGC6851&3.62071&-24.23&17.44&1.677&1.28&0.41&0.83&0.94&3.69& 0 & 3.9 & \nodata & \nodata & \nodata &1\\
NGC6868&4.27965&-25.53&19.43&5.704&2.82&0.58&0.93&1.04&4.09& 0.1 & 5.8 & 0.22 & 9.2 & 5 &1\\
NGC687&3.93865&-24.86&18.40&3.440&2.67&0.56&0.93&1.04&4.02& 0 & 5.8 & \nodata & \nodata & \nodata &1\\
NGC6958&3.65829&-24.28&17.58&2.550&1.42&0.43&0.84&0.94&3.69& 0 & 4 & 0.28 & 3 & 5 &1\\
NGC7014&3.97023&-24.88&18.54&4.332&2.38&0.52&0.91&1.02&3.96& 0.1 & 5.4 & \nodata & \nodata & \nodata &1\\
NGC7029&3.66291&-24.24&17.65&2.197&1.08&0.36&0.83&0.93&3.68& 0 & 3.9 & \nodata & \nodata & \nodata &1\\
NGC7144&3.65758&-23.92&17.93&2.500&1.27&0.41&0.88&0.99&3.62& -0.1 & 4.7 & 0.22 & 3.8 & 3 &1\\
NGC7145&3.20764&-23.32&16.29&0.791&1.17&0.39&0.85&0.95&3.44& -0.1 & 4.2 & \nodata & \nodata & \nodata &1\\
NGC7173&3.40097&-23.46&17.11&2.789&1.58&0.45&0.87&0.97&3.81& -0.2 & 4.6 & \nodata & \nodata & \nodata &1\\
NGC7192&3.74799&-24.40&17.91&1.982&1.06&0.35&0.89&0.99&3.54& 0 & 4.9 & 0.30 & 5.7 & 5 &1\\
NGC7196&3.76349&-24.59&17.79&3.842&1.76&0.46&0.89&1.00&3.98& 0 & 4.9 & \nodata & \nodata & \nodata &1\\
NGC7200&3.30108&-22.88&17.19&3.205&2.20&0.50&0.73&0.81&3.51& -0.2 & 3.4 & \nodata & \nodata & \nodata &1\\
NGC7236&3.98476&-25.19&18.30&1.370&1.99&0.48&0.88&0.99&4.34& 0 & 4.7 & \nodata & \nodata & \nodata &1\\
NGC7332&3.06963&-22.24&16.68&1.632&2.32&0.52&0.91&1.02&3.79& 0 & 5.4 & -0.02 & 3.7 & 5 &1\\
NGC7562&3.94391&-24.91&18.38&3.361&2.61&0.55&0.94&1.05&3.84& 0.1 & 6 & \nodata & \nodata & \nodata &1\\
NGC7768&4.33595&-25.82&19.43&3.583&2.02&0.49&0.82&0.92&3.67& 0.2 & 2.7 & \nodata & \nodata & \nodata &1\\
NGC80&4.18289&-25.41&19.07&5.505&3.11&0.60&0.91&1.02&4.08& 0.1 & 5.4 & \nodata & \nodata & \nodata &1\\
E376-G007&3.73038&-24.36&17.86&2.595&1.50&0.44&0.84&0.94&3.96& 0 & 4 & \nodata & \nodata & \nodata &1\\
IC3370&4.11142&-25.37&18.76&2.350&1.01&0.34&0.85&0.95&3.72& 0.2 & 4.2 & \nodata & \nodata & \nodata &1\\
IC4943&3.36849&-23.06&17.35&2.325&1.39&0.43&0.87&0.98&3.56& -0.2 & 4.6 & \nodata & \nodata & \nodata &1\\
NGC194&3.94334&-24.92&18.37&2.461&2.88&0.58&0.91&1.02&3.73& 0.1 & 5.4 & \nodata & \nodata & \nodata &1\\
NGC227&3.98541&-25.07&18.42&3.893&1.57&0.45&0.88&0.98&3.83& 0.1 & 4.7 & \nodata & \nodata & \nodata &1\\
NGC2300&3.75165&-24.42&17.91&4.179&3.14&0.61&0.97&1.08&4.13& -0.1 & 6.8: & 0.20 & 11.7 & 6 &1\\
NGC2832&4.61158&-26.27&20.36&9.915&2.44&0.53&0.92&1.03&3.93& 0.3 & 4.0: & 0.30 &10.0& 4 &1\\
NGC2865&3.83118&-24.57&18.15&1.704&1.04&0.35&0.80&0.90&3.61& 0.1 & 3.4 & \nodata & \nodata & \nodata &1\\
NGC4373&4.33406&-25.87&19.37&3.182&1.11&0.37&0.86&0.97&3.90& 0.2 & 3.2 & \nodata & \nodata & \nodata &1\\
NGC4616&3.76140&-24.44&17.93&1.653&1.86&0.47&0.90&1.01&4.12& 0 & 5.1 & \nodata & \nodata & \nodata &1\\
NGC4692&4.24744&-25.52&19.29&4.899&3.40&0.62&0.96&1.08&3.79& 0.2 & 6.5 & 0.30 &10.0& 4 &1\\
NGC4839&4.34453&-25.90&19.39&3.881&2.75&0.57&0.88&0.99&3.72& 0.2 & 3.4 & \nodata & \nodata & \nodata &1\\
NGC4915&3.78878&-24.55&17.96&2.436&1.34&0.42&0.83&0.93&3.96& 0 & 3.9 & \nodata & \nodata & \nodata &1\\
NGC4946&3.78252&-24.40&18.08&2.400&2.74&0.57&0.91&1.02&3.99& 0 & 5.4 & \nodata & \nodata & \nodata &1\\
NGC6051&4.34637&-26.12&19.18&2.805&1.94&0.48&0.99&1.11&4.25& 0.2 & 5.2 & \nodata & \nodata & \nodata &1\\
NGC7192&3.74799&-24.40&17.91&1.982&1.06&0.35&0.89&0.99&3.54& 0 & 4.9 & 0.30 & 5.7 & 5 &1\\
NGC1387&3.54978&-23.66&17.66&1.496&2.09&0.49&0.96&1.08&4.27& -0.2 & 8.9 & \nodata & \nodata & \nodata &1\\
NGC1399&3.92221&-24.99&18.19&5.903&2.11&0.50&0.93&1.05&4.05& 0.1 & 5.8 & 0.33 & 5.1 & 3 &1\\
VCC1036&3.07700&-20.74&18.21&1.231&0.94&0.40&0.86&0.96&3.11& -0.5 & 5.9 & -0.50 & 3.5 & 1,7 &5\\
VCC1073&3.03600&-20.53&18.22&1.135&1.52&\nodata&\nodata&\nodata&3.24& -0.5 & 5.3 & \nodata & \nodata & \nodata &5\\
VCC1308&3.09100&-18.51&20.51&6.153&0.31&0.28&\nodata&\nodata&2.36& -0.7 & 3.4 & -0.46 & 3.6 & 7 &5\\
VCC1488&2.92300&-19.20&18.98&1.322&0.39&\nodata&\nodata&\nodata&2.58& -0.6 & 3.7 & \nodata & \nodata & \nodata &5\\
VCC452&3.02300&-18.02&20.66&1.978&0.17&0.18&\nodata&\nodata&1.92& -0.7 & 3.3 & \nodata & \nodata & \nodata &5\\
VCC745&2.93900&-19.99&18.27&2.763&0.67&\nodata&\nodata&\nodata&2.96& -0.6 & 3.9 & \nodata & \nodata & \nodata &5\\
VCC917&2.88600&-19.01&18.99&1.901&0.41&\nodata&\nodata&\nodata&2.66& -0.6 & 3.4 & -0.62 & 4.4 & 7 &5\\
VCC940&3.11300&-20.01&19.12&1.109&0.53&0.31&\nodata&\nodata&2.82& -0.6 & 3.7 & -0.70 & \nodata & 1 &5\\
IC4011&3.38804&-22.93&16.11&1.565&2.40&0.41&0.90&1.01&3.80& -0.2 & 7 & \nodata & \nodata & \nodata &1\\
IC3960&3.45842&-23.37&15.84&2.351&3.05&0.46&0.95&1.07&4.25& -0.2 & 8.6 & \nodata & \nodata & \nodata &1\\
IC3973&3.30529&-23.09&16.65&3.280&5.07&0.50&0.93&1.04&4.17& -0.3 & 7.9 & 0.13 & 9.8 & 4 &1\\
IC4021&3.29219&-22.80&15.32&2.283&3.10&0.46&0.94&1.06&3.79& -0.3 & 8.2 & \nodata & \nodata & \nodata &1\\
IC3393&2.91800&-18.46&19.70&1.983&0.75&\nodata&\nodata&\nodata&3.02& -0.7 & 4 & \nodata & \nodata & \nodata &5\\
NGC4431&3.17700&-20.88&18.57&1.489&2.34&0.54&0.79&0.88&3.36& -0.5 & 4.4: & -0.30 &10.0& 4 &5\\
UGC7436&2.99300&-19.94&18.60&1.649&1.73&\nodata&\nodata&\nodata&3.31& -0.6 & 3.7 & \nodata & \nodata & \nodata &5\\
IC0225&3.06400&-19.27&19.62&0.906&0.50&\nodata&\nodata&\nodata&2.79& -0.66 & 1.6 & \nodata & \nodata & \nodata &5\\
IC3328&3.13500&-20.05&19.19&1.335&0.81&\nodata&\nodata&\nodata&3.06& -0.6 & 4.1 & \nodata & \nodata & \nodata &5\\
IC3381&2.98000&-19.65&18.82&2.083&1.39&\nodata&\nodata&\nodata&3.21& -0.6 & 5 & \nodata & \nodata & \nodata &5\\
IC3461&2.98900&-18.93&19.59&4.193&0.36&0.30&\nodata&\nodata&2.48& -0.7 & 3.6 & \nodata & \nodata & \nodata &5\\
IC3468&3.17500&-21.22&18.22&0.431&1.94&\nodata&\nodata&\nodata&3.34& -0.4 & 6 & \nodata & \nodata & \nodata &5\\
IC3653&2.85500&-19.90&17.94&1.381&1.27&\nodata&\nodata&\nodata&3.18& -0.6 & 4.8 & \nodata & \nodata & \nodata &5\\
NGC4121&2.94500&-20.93&17.36&1.570&1.03&0.47&0.88&0.98&3.13& -0.4 & 6.4 & \nodata & \nodata & \nodata &5\\
NGC4308&2.86500&-19.76&18.13&2.522&3.96&0.27&0.83&0.93&3.40& -0.7 & 6.5 & \nodata & \nodata & \nodata &5\\
NGC4328&2.93100&-18.75&19.48&2.519&0.40&\nodata&\nodata&\nodata&2.63& -0.7 & 3.7 & \nodata & \nodata & \nodata &5\\
NGC4415&3.12900&-20.98&18.24&0.619&0.73&\nodata&\nodata&\nodata&3.01& -0.4 & 4.0: & -0.20 &10.0& 4 &5\\
PGC41682&3.09100&-20.48&18.54&1.127&0.68&0.30&\nodata&\nodata&2.97& -0.5 & 3.9 & \nodata & \nodata & \nodata &5\\
IC3081&3.13700&-18.86&20.39&10.923&2.37&\nodata&\nodata&\nodata&1.71& -0.6 & 7.3 & \nodata & 1.71 & \nodata &5\\
IC3344&2.96700&-19.01&19.39&4.841&0.29&\nodata&\nodata&\nodata&2.26& -0.7 & 3.7 & \nodata & \nodata & \nodata &5\\
IC3735&3.43200&-21.03&19.70&2.022&0.30&\nodata&\nodata&\nodata&2.33& -0.5 & 3.4 & \nodata & \nodata & \nodata &5\\
IC3779&3.36200&-18.72&21.66&13.792&2.10&\nodata&\nodata&\nodata&3.35& -0.6 & 6.2 & \nodata & \nodata & \nodata &5\\
\enddata
\tablerefs{
1 = Jerjen et al. (2004);
2 = Proctor et al. (2004);
3 = Li et al. (2007);
4 = Sanchez-Blazquez et al. (2006)
5 = Annibali et al. (2007)
6 = Serra et al. (2008)
7 = Chilingarian (2009);
8 = Paudel et al. (2010);
9 = Harris (2003);
10 = De Angeli (2005);
11 =  Marin-Franch et al.(2009);
12 = Forbes\& Bridges(2010);
13 = Harris et al. 1997;
14 = Caldwell et al. 2011;
15 = Schweizer\& Seitzer 2007;
16 = Evstigneeva et al. 2007;
17 = Hasegan et al. 2005;
18 = Mateo 1998;
19 = Durrell et al.1996;
20 = Chiligarian \& Mamon 2008;
21 = Chilingarian et al. 2011;
22 = Mieske et al. 2008;
23 = Monachesi et al. 2011;
24 = Dabringhausen et al 2008
}
\end{deluxetable}

\clearpage

\begin{deluxetable}{lcc}
\tabletypesize{\scriptsize} \tablecaption{Gap values $u_k$ for different values of
the number $k$ of clusters.\label{tbl_gap}} \tablewidth{0pt} \tablehead{
\colhead{Number of clusters} & \colhead{Gap}&\colhead{$u_k$} }
\startdata
k=1&1.658168&-0.3570766\\
k=2&1.592197&-0.1841725\\
k=3&1.804713&-0.7284219\\
k=4&2.592263&0.0873069\\
k=5&2.529849&-0.5264599\\
k=6&2.43953&0.0961393\\
k=7&2.356531&0.08300998\\
k=8&2.572305&-0.4335349\\
\hline
\enddata
\end{deluxetable}

\begin{deluxetable}{lcccc}
\tabletypesize{\scriptsize} \tablecaption{Average properties with
standard errors of the four main groups.
\label{tbl_meandata}} \tablewidth{0pt} \tablehead{
\colhead{Groups} & \colhead{FK1}&\colhead{FK3} & \colhead{FK5} &
\colhead{FK6}} \startdata
Number of members & 210& 57& 39& 77\\
$log\sigma_0 (kms^{-1})$ & 2.3385 $\pm$ 0.0076 & 1.3134 $\pm$
0.0329 & 1.6088 $\pm$ 0.0214 & 0.9679 $\pm$ 0.0346 \\
$logR_h (pc)$ & 3.8646$\pm$ 0.0228 & 0.7363 $\pm$ 0.0537 &
3.013 $\pm$ 0.0333 & 0.6397 $\pm$ 0.0454 \\
$M_K (mag)$ & -24.691 $\pm$ 0.058 & -13.305 $\pm$ 0.214 & -19.776
$\pm$ 0.143 & -11.229 $\pm$ 0.198 \\
$\mu_{h,K} (mag \  arcsec^{-2}$) & 18.199 $\pm$ 0.059 & 13.944
$\pm$ 0.169 & 18.857 $\pm$ 0.176 & 15.537 $\pm$ 0.177 \\
$M_{vir}/L_K (M_{vir,\odot}/L_{K,\odot}$) & 3.171 $\pm$ 0.116 &
1.772 $\pm$ 0.239 & 2.489 $\pm$ 0.421 & 1.750 $\pm$ 0.147 \\
$ M/L_V (M_{\odot}/L_{V,\odot})$& 1.9073 $\pm$ 0.049 & 3.173
$\pm$ 0.290 & 1.182 $\pm$ 0.170 & 2.716 $\pm$ 0.230 \\
log($M_{vir}) (M_{\odot})$& 11.625 $\pm$ 0.034 & 6.753 $\pm$ 0.106
& 9.6175 $\pm$ 0.0957 & 5.9656 $\pm$ 0.0945 \\
$ U-B (mag)$ & 0.466 $\pm$0.00541 & 0.3205 $\pm$ 0.0419 & 0.3575
$\pm$ 0.0378 & 0.0671 $\pm$ 0.0150 \\
$ B-V (mag) $ & 0.893 $\pm$ 0.003 & 0.793 $\pm$ 0.032 & 0.870
$\pm$ 0.034 & 0.689 $\pm$ 0.01 \\
$V-I (mag)$ & 1.0001 $\pm$ 0.0034 & 1.0527 $\pm$ 0.027 & 0.9375
$\pm$ 0.022 & 0.889 $\pm$ 0.0078 \\
$B-K (mag) $ & 3.818 $\pm$ 0.017 & 3.732 $\pm$ 0.270 & 2.903 $\pm$
0.066 & 3.299 $\pm$ 0.100 \\
$[Fe/H] (dex)$ & 0.04495 $\pm$ 0.0087 & -0.5472 $\pm$ 0.038 &
-0.6069 $\pm$ 0.021 & -1.5030 $\pm$ 0.380 \\
Age (Gyr) & 5.904 $\pm$ 0.159 & 10.637 $\pm$ 0.468 & 4.817 $\pm$
0.646 & 10.217 $\pm$ 0.226 \\
\hline
\enddata
\end{deluxetable}

\begin{deluxetable}{lcccc}
\tabletypesize{\scriptsize} \tablecaption{Average properties with
standard errors of three subsamples taken from \citet{for08} and
from our smaller sample. \label{tbl_bias}} \tablewidth{0pt}
\tablehead{ \colhead{Samples} &
\colhead{$M_K$(F08)}&\colhead{$M_K$(ours)} & \colhead{$\sigma _0$
(F08)} & \colhead{$\sigma _0$ (Ours)}} \startdata 1 (GCs)&
-11.327$\pm$ 1.87 & -11.339$\pm$0.156 & 12.4$\pm$0.841 &
12.077$\pm$0.686\\
2 (IMOs) & -13.98$\pm$0.291& -14.145$\pm$0.246& 28.58$\pm$2.45 &
29.13 $\pm$ 2.44 \\
3 (Es) & -23.67 $\pm$ 0.084 & -23.943 $\pm$0.124 & 185.5 $\pm$3.84

& 196.81 $\pm$ 5.31 \\
\hline
\enddata
\end{deluxetable}

\begin{deluxetable}{llll}
\tabletypesize{\scriptsize} \tablecaption{Robust multivariate regression on
the four main groups in $k1$-$k3$ space.\label{tbl_robustfit}}
\tablewidth{0pt} \tablehead{\colhead{Group}&\colhead{$a$}&\colhead{$b$}&\colhead{rms}}
\startdata
FK1&-0.567413&0.314669&0.06823\\
FK3&0.238372&0.244655&0.09654\\
FK5&0.133827&0.141191&0.17752\\
FK6&0.378629&0.155797&0.11949\\
\hline
\enddata
\end{deluxetable}

\end{document}